# Near-room-temperature magnetoelectric coupling engineered through inversion-breaking tilts in a bulk perovskite polytype

*Struan Simpson*[1]*, *Urmimala Dey*[2], *Martin R. Lees*[3], *Ivan Da Silva*[4], *Nicholas C. Bristowe*[5], *and Mark S. Senn*[1]*

[1] Department of Chemistry, University of Warwick, Gibbet Hill, Coventry, CV4 7AL, U.K.

[2] Luxembourg Institute of Science and Technology, Avenue des Hauts-Fourneaux 5, L4362, Esch-sur-Alzette, Luxembourg

[3] Department of Physics, University of Warwick, Gibbet Hill, Coventry, CV4 7AL, U.K.

[4] ISIS Facility, Rutherford Appleton Laboratory, Harwell Oxford, Didcot OX11 0QX, U.K.

[5] Centre for Materials Physics, Durham University, South Road, Durham, DH1 3LE, U.K.

ABSTRACT. Systematic strategies to design properties such as ferroelectricity or magnetoelectric coupling are well established in simple perovskite materials, but they remain scarce in more complex framework structures. Using a hexagonal polytype of the ternary manganite $A$MnO$_3$ ($A$ = Ba, Sr, Ca) as a model system, we introduce a symmetry-guided design principle in which an inversion-breaking rigid-unit mode (RUM) serves as a single structural instability generating both polar and ferromagnetic orders within a bulk material. Symmetry analysis and first-principles calculations reveal that co-operative tilts of the $Mn_2O_9$ bioctahedral dimers generate both a spontaneous polarization and a ferromagnetic moment. High-resolution diffraction and magnetic susceptibility measurements show the structural and magnetic orders persist as high as 450 K and 280 K, respectively, highlighting the untapped potential of framework structures which deviate from simple perovskite motifs to be designed to host useful ferroic properties. Our approach establishes a transferable symmetry-based framework to engineer ferroelectric and magnetoelectric states across chemically diverse framework architectures.

INTRODUCTION. The rapid expansion of artificial intelligence has intensified the demand for energy-efficient memory and logic technologies. Addressing this demand requires materials that enable low-power control of magnetic states, motivating renewed interest in magnetoelectric (ME) multiferroics in which magnetic order can be manipulated via applied electric fields.[1] Prototype ME switching devices have demonstrated substantial improvements in endurance and energy efficiency compared to existing devices.[2] Nevertheless, materials with robust ME coupling near room temperature remain exceptionally rare. This scarcity stems from the longstanding challenge of combining the chemical ingredients required for large ferroelectric (FE) polarizations (e.g., $d^0$ cations such as $Ti^{4+}$) and ferromagnetism (i.e., cations with unpaired electrons) within the same material.[3] The rational design of new ME materials represents an enduring challenge in modern materials chemistry.

Recently, substantial progress has been made in the design of new ME materials through group-theoretical methods. The fundamental principle of this approach is to identify a material system in which a ferroelectric (FE) structural distortion and weak ferromagnetism (wFM) derive from a common symmetry-breaking origin, enabling both properties to be controlled via an intermediary distortion. This principle was first elucidated in the form of the hybrid-improper ferroelectric mechanism, where two non-polar octahedral tilt modes may be combined to generate and control a spontaneous polarization.[4] This was first demonstrated in $PbTiO_3/SrTiO_3$ superlattices [5] before being extended to a swathe of layered perovskite materials such as $A_3B_2O_7$ Ruddlesden-Popper phases [6–8] and $AA'B_2O_7$ Dion-Jacobson phases.[9,10] Importantly, this purely geometric source of ferroelectricity is compatible with the presence of unpaired electrons on the *B* site, enabling the coexistence of polar and magnetic orders within the same material. This approach has since been extended to enumerate robust mechanisms of ME coupling in other perovskite-based materials, where specific combinations of symmetry-breaking distortions can be designed that yield directly coupled FE and wFM orders.[11–13] Nevertheless, magnetic ordering in ME perovskite-based materials is often limited to cryogenic temperatures.[14–18] This has inevitably limited the practicality of these systems for functional applications, spurring ongoing efforts to discover new ME materials with more practical operating temperatures.

One possible workaround is to apply group-theoretical approaches to structures that deviate from simple corner-sharing connectivity between their polyhedra. Frameworks incorporating edge- or face-sharing polyhedra offer alternative exchange pathways characterized by strong orbital overlap between nearest-neighbor metal sites, hence their enhanced exchange interactions could lead to higher magnetic ordering temperatures.[19–22] However, these frameworks are generally thought to be less flexible than those comprised of corner-sharing polyhedra [23], so they are often assumed to lack practical structural instabilities that can be readily induced to achieve coupled polar and magnetic orders. Indeed, this has represented the key bottleneck inhibiting the group-theoretical design of ferroic functionalities in framework structures which deviate from conventional perovskite motifs.

Here, we exploit a newly discovered rigid-unit mode (RUM) in a four-layer hexagonal (4H) polytype of the perovskite structure (Fig. 1a) to induce mutually coupled polar and wFM orders close to room temperature (~280 K, ~7 °C). Perovskite polytypes (sometimes referred to as "hexagonal perovskites") are generally characterized by face-sharing connectivity between their octahedra, hence they provide an ideal platform to identify new forms of structural instability that could yield useful functionalities in more complex framework topologies. Using 4H-$A$MnO$_3$ ($A$ = $Ba^{2+}$, $Sr^{2+}$, $Ca^{2+}$) as a model system (Fig. 1b), we use high-resolution diffraction, first-principles calculations, and magnetic susceptibility measurements to show how a single inversion-breaking RUM can be engineered in this system to induce the mutually coupled polar and wFM orders required for ME coupling. In contrast to established tilt-driven mechanisms in perovskite-based materials,[11–13] the mechanism we identify here is remarkably simple, requiring minimal symmetry-breaking distortions to fulfil the necessary chemical ingredients for ME coupling. More generally, our insights showcase the potential of stabilizing inversion-breaking RUMs in more complex framework topologies to confer them with newfound functionality.

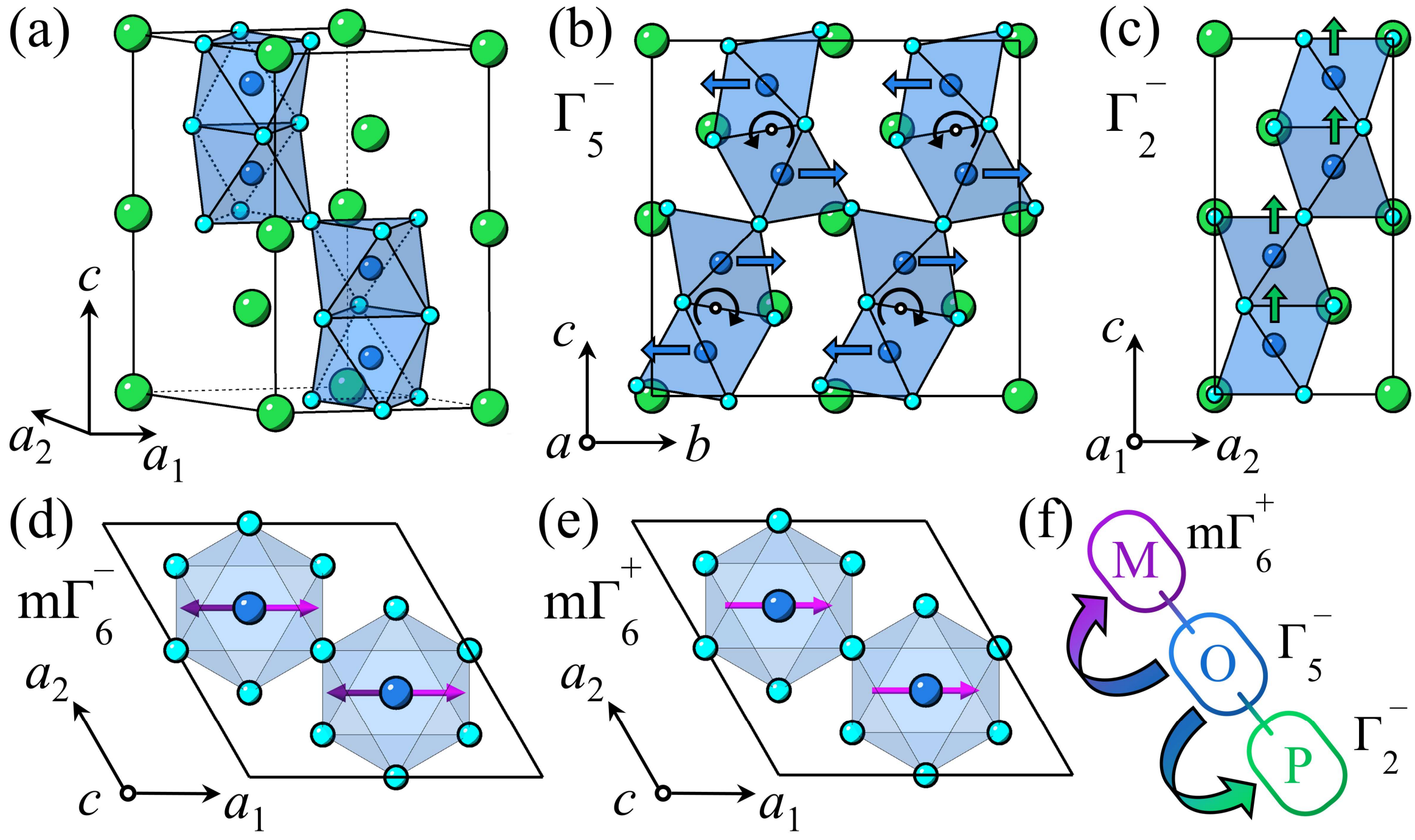

**Figure 1.** Symmetry-breaking distortions in 4H-$SrMnO_3$. (a) The $P6_3/mmc$ aristotype structure, with green, blue, and cyan atoms representing the Sr, Mn, and O sites, respectively. (b) The room-temperature $Cmc2_1$ structure, highlighting the $\Gamma_5^-$ tilt mode of the $Mn_2O_9$ bioctahedra (black arrows) and the accompanying antipolar Mn displacements (blue arrows). (c) The $\Gamma_2^-$ polar distortion depicted with respect to the parent structure; only the Mn displacements (green arrows) have been shown for visual clarity. (d), (e) The AFM and wFM spin components, transforming as $m\Gamma_6^-$ and $m\Gamma_6^+$, respectively. For the $m\Gamma_6^-$ component, light and dark purple arrows map to the Mn sites at $z \approx 3/8$, 7/8 and $z \approx 1/8$, 5/8, respectively. (f) The fundamental design principle yielding ME coupling in 4H-$SrMnO_3$, showing how octahedral tilting (O) simultaneously induces both an electrical polarization (P) and a spontaneous magnetization (M).

MATERIALS AND METHODS. First-principles calculations were performed within the density functional theory (DFT) framework as implemented in the Vienna Ab initio Simulation Package (VASP), version 6.4.3.[24] Projector augmented-wave (PAW) pseudopotentials (PBE, version 5.4) [25] were used with the following valence electron configurations: $4s^2 4p^6 5s^2$ (Sr), $3p^6 3d^5 4s^2$ (Mn), and $2s^2 2p^4$ (O). The PBEsol flavor of general gradient approximation (GGA) exchange-correlation functional [26] was employed to accurately describe the structural properties of bulk 4H-$SrMnO_3$. The energy convergence criterion was set at $10^{-9}$ eV for all calculations, and full structural relaxations were carried out until the Hellmann–Feynman forces on each atom were less than 1 meV/Å. Correlation effects were treated within the GGA+$U$ formalism introduced by Dudarev et al. [27] using an effective on-site Hubbard parameter of $U_{\mathrm{eff}} = 4.5$ eV for the Mn 3d orbitals.[28,29] We varied the $U_{\mathrm{eff}}$ values within a reasonable range to examine the stability of the ground state magnetic order. We performed convergence tests for the $P6_3/mmc$ aristotype structure with the ground state A-type antiferromagnetic (AFM) spin configuration and found that a 700 eV plane wave energy cutoff and a 5 x 5 x 3 **k**-point mesh were sufficient to converge the energy differences, atomic forces, and stresses within 0.01 meV/f.u., 0.2 meV/Å, and 0.2 GPa, respectively. Spin-orbit coupling (SOC) effects were included self-consistently in the noncollinear calculations. The Berry phase method [30,31] within VASP is used to compute spontaneous electric polarization. Phonon calculations were performed using the finite displacement method for a 2 x 2 x 2 phonon supercell, and phonon dispersions were generated with PHONOPY.[32] The ISOTROPY software suite was employed for subsequent symmetry analysis.[33]

Pristine polycrystalline samples of 4H-$A$$MnO_3$ ($A = Ba^{2+}, Sr^{2+}, Ca^{2+}$) were synthesized according to a previous method.[34] Stoichiometric quantities of $BaCO_3$ (Sigma-Aldrich, ≥99.9%), $SrCO_3$ (Sigma-Aldrich, ≥99.9%), $CaCO_3$ (Sigma-Aldrich, ≥99.9%), and $MnO_2$ (Sigma-Aldrich, ≥99%) were mixed and ground using an agate mortar and pestle before being pressed into 10 mm pellets and heated to 900 °C in an alumina crucible for 10 h. The pellets were then reground and reheated

twice for 10 h at 1250 °C h to ensure both high purity and crystallinity. Due to the tendency for $SrMnO_3$ to reduce above ~1050 °C,[34] each sample was also subjected to an additional 24 h dwell at 600 °C to ensure the nominal oxygen stoichiometry was accurate. This was later corroborated by Rietveld refinements against neutron powder diffraction (NPD) data, which showed the Sr, Mn, and O sites could each be modelled to within ±1% of their nominal occupancies. Laboratory X-ray diffraction (XRD) measurements showed that the samples prepared this way typically contained only trace impurities of a $Mn_3O_4$ spinel phase (<0.5 wt.%).

High-resolution synchrotron XRD (S-XRD) measurements [35] were performed at the ID22 beamline (ESRF, France) using a 35 keV X-ray beam and a 13-channel Si (111) multi-analyzer stage. The sample was loaded into a 0.5 mm borosilicate capillary before being measured in the temperature range 10–500 K for ~45 minutes at each temperature point. Additional S-XRD measurements were performed at the I11 beamline (Diamond Light Source, U.K.) using a 15 keV X-ray beam. The sample was loaded into a 0.3 mm quartz capillary to minimize absorption effects, and data were collected continuously in the ranges 100–300 K and 25–300 °C using a position-sensitive Mythen detector. NPD measurements were performed on the GEM diffractometer (ISIS Neutron and Muon Source, U.K.).[36] ~1.5 g of 4H-$SrMnO_3$. were loaded into a vanadium can for these measurements. Data were collected for ~3 h at 10 K, but due to time limitations data were collected for ~1.5 h at 290 K. Symmetry-adapted Rietveld fits were performed with Topas using the jEdit interface.[37,38] Full crystallographic data from the refinements have been provided in the Supplementary Information. Additional symmetry analysis was performed using the ISODISTORT,[39,40] INVARIANTS,[41,42] and ISOTILT [43,44] webtools. Mode amplitudes have been normalized with respect to the parent cell, but note since all of the symmetry-breaking distortions transform as Γ-point irreps, these are the same as the subcell-normalized amplitudes.

DC magnetic susceptibility data were collected using a Quantum Design Magnetic Properties Measurement System (MPMS). Zero-field-cooled (ZFC) and field-cooled (FC) data were collected between 5 and 300 K using a 1000 Oe field. Hysteresis loops were recorded at various temperatures between 200 K and 300 K and fields between –50 and 50 kOe. DC resistivity data were collected in the range 250–400 K with a Quantum Design Physical Property Measurement System (PPMS). A four-probe method was used to measure polycrystalline bars with typical dimensions of 5 mm x 3 mm x 1 mm.

THEORETICAL BASIS. The basis of our design philosophy is informed by our recent study of an exotic polar instability in the six-layer hexagonal (6H) polytype of the canonical ferroelectric perovskite $BaTiO_3$,[45] which provides a prototype for the symmetry-coupled polar distortions we seek to generalize here. In 6H-$BaTiO_3$, FE order emerges due to a second-order Jahn-Teller (SOJT) distortion associated with the $Ti^{4+}$ ($d^0$) cations. This distortion primarily manifests in terms of antipolar displacements of the $Ti^{4+}$ cations within the hexagonal basal plane, but because the $TiO_6$ octahedral faces are tilted by ~9° out of the plane, a secondary polar component is induced along the *c* axis to ensure the $Ti^{4+}$ cations displace towards the octahedral face centers. Symmetry analysis shows the coupled antipolar and polar displacements transform with respect to the $P6_3/mmc$ aristotype structure as the $\Gamma_5^-$ and $\Gamma_2^-$ irreducible representations (irreps), respectively. We further predicted that analogous distortions could emerge in several related polytypes (i.e., 4H, 8H, and 12H) that share the same $P6_3/mmc$ aristotype symmetry. However, so far only the 6H polytype of $BaTiO_3$ has been experimentally realized. Additionally, the polarization is contingent on the SOJT distortion of the $Ti^{4+}$ cations, so this mechanism precludes coupling to magnetic order. These limitations motivated us to search for alternative structural instabilities that could yield analogous polarization coupling while remaining chemically compatible with magnetism.

Recently, we predicted a class of inversion-breaking RUMs [46] in several perovskite polytypes that are capable of inducing a spontaneous polarization on an improper basis. Formally, RUMs represent low-energy phonon modes in framework materials that do not distort the internal bonding environments of the coordination polyhedra,[47] hence they are expected to be lower in energy than all of the other vibrational modes of the framework. RUMs thus represent feasible structural deformations which are the most likely to have sufficiently low frequencies to "freeze out" with respect to the undistorted parent structure. For example, in conventional (3C) $ABX_3$ perovskites, RUMs are more familiarly known in terms of co-operative tilts or rotations of the $BX_6$ octahedra which can be frozen out with respect to the $Pm\bar{3}m$ aristotype structure to yield lower-symmetry structures.[48,49] These tilts can be controlled by varying the degree of size mismatch between the average radii of the $A$- and $B$-site cations,[50] and this represents a more general strategy by which RUMs may be manipulated to induce phase transitions in other materials as well.

Beyond their initial significance in characterizing distorted structures, RUMs have important implications for material functionality. In particular, they are foundational to the mechanisms of hybrid-improper ferroelectricity and ME coupling which have been elucidated in perovskite-based materials.[4,12] In contrast, RUMs in edge- or face-sharing polyhedral frameworks have historically been considered unfavorable due to the higher number of geometric constraints imposed upon the framework, inhibiting their flexibility.[49] This is precisely what has previously prohibited scope for the rational design of functional properties in more complex framework structures. However, our recent symmetry insights in perovskite polytype structures have overturned this longstanding consensus: several polytypes host RUMs involving intrinsic antipolar displacements of the $B$-site cations,[46] thereby breaking inversion symmetry directly without the need for extraneous symmetry-breaking distortions (e.g., cation/anion ordering or SOJT distortions). This contrasts

with 3C perovskites, where additional symmetry-breaking mechanisms such as cation order or SOJT distortions are required to couple to the octahedral tilts to induce FE order.[12] Consequently, hexagonal polytypes offer appealing advantages compared to 3C perovskites as a platform to design polar or ME functionalities.

To demonstrate this design principle, we consider the 4H perovskite polytype (Fig. 1a). In this structure type, the $\Gamma_5^-$ RUM corresponds to an in-phase tilting of the $B_2X_9$ bioctahedra about an axis lying in the hexagonal basal plane (Fig. S1). Condensing the action of the associated order parameter on the high-symmetry parent phase yields a series of structures with either $C222_1$, $Cmc2_1$, or $P2_1$ symmetry, depending on the orientation of the tilt axis (i.e., order parameter direction). While all three structures belong to piezoelectric point groups, the $Cmc2_1$ (Fig. 1b) and $P2_1$ structures (Fig. S1c) are also polar point groups ($mm2$ and 2, respectively), therefore they may display a spontaneous polarization. Microscopically, this is due to a series of out-of-plane displacements of the $A$, $B$, and $X$ atoms transforming as the $\Gamma_2^-$ distortion modes (Fig. 1c), demonstrating the same polarization coupling that we identified in 6H-$BaTiO_3$.[45] By analogy to how octahedral tilt modes are stabilized in 3C perovskites,[50] $A$/$B$-site size mismatch is expected to condense the $\Gamma_5^-$ RUM, thus inducing static polyhedral tilting and a spontaneous polarization. Crucially, this RUM-based mechanism is fully compatible with the presence of magnetic $B$-site cations, providing a viable route to combine FE and wFM orders within a single phase. In the following, we demonstrate the experimental realization of our design strategy.

RESULTS AND DISCUSSION. To demonstrate how FE order may be induced via the $\Gamma_5^-$ mode, we consider the ternary perovskite system 4H-$A$$MnO_3$ ($A$ = $Ba^{2+}$, $Sr^{2+}$, $Ca^{2+}$). The magnetic $Mn^{4+}$ ($d^3$) cations make this system an ideal platform to assess the compatibility of the induced FE and magnetic orders, and this system is generally characterized by relatively high Néel temperatures ($T_N \approx 263$ K and 280 K for $A = Ba^{2+}$ and $Sr^{2+}$, respectively [51,52]). For $A = Ba^{2+}$, the crystal structure retains the aristotype $P6_3/mmc$ symmetry down to at least 5 K;[53] for $A = Sr^{2+}$, a structural phase transition was previously reported to occur below ~350 K.[52] The low-symmetry 4H-$SrMnO_3$ structure was previously assigned to the noncentrosymmetric (but nonpolar) $C222_1$ space group, which is consistent with the predictions of our RUM analysis. However, our high-resolution S-XRD data (Fig. 2a) reveal that this model provides a poor description of the observed superstructure reflections (Figs. S2, S3), and Rietveld refinements yielded consistently worse fitting statistics compared to the alternative $Cmc2_1$ or $P2_1$ models (Table S1). Furthermore, variable-temperature S-XRD measurements (Fig. 2b) indicate that the structural transition temperature ($T_s$) is closer to 450 K, suggesting previous studies underestimated $T_S$ by ~100 K. These discrepancies motivated us to reinvestigate the structure of 4H-$SrMnO_3$ to determine the correct tilt orientation and establish the possibility of a RUM-induced FE order.

Exploiting the strong sensitivity of powder diffraction to lattice strain, we used the coupling between the ferroelastic strain and the $\Gamma_5^-$ tilt orientation to distinguish between candidate space group assignments (see Supplementary Note 2). We performed significance tests with respect to the orientation of the $\Gamma_5^-$ tilt axis ($\varphi(\Gamma_5^-)$) against our high-resolution S-XRD data collected on ID22 (Fig. 2c). At 10 K, we find that a $Cmc2_1$ structural model ($\varphi = 30°$) leads to a significantly better fit ($R_{wp} = 9.3\%$) compared to the canonical $C222_1$ model ($\varphi = 60°$; $R_{wp} = 10.2\%$), despite both models employing the same number of refined parameters. Additionally, the $C222_1$ model predicts a subtle splitting of the superstructure reflections (Fig. S3) which is not observed within the very high instrumental resolution of our S-XRD measurements ($\Delta d/d \approx 1 \times 10^{-4}$), whereas the $Cmc2_1$ model accurately reproduces their appearance. This assignment is further supported by our DFT calculations, which confirm that the ground-state structure is described by $Cmc2_1$ symmetry (**Figure 3**). Although the reduction in ferroelastic strain above 100 K limits our sensitivity to the space group assignment at higher temperatures, the absence of any anomalies in the lattice parameters (Fig. S4, Fig. S9) supports a single low-symmetry phase below $T_S \approx 450$ K. Notably, this contrasts with 6H-$BaTiO_3$, where a $Cmc2_1 \rightarrow C222_1$ transition at $T_c \approx 70$ K is readily distinguished by discontinuities in the lattice parameters.[45] Therefore, we conclude that 4H-$SrMnO_3$ adopts the polar $Cmc2_1$ structure for all $T < T_S$. Refined structural parameters for the final $Cmc2_1$ model are provided in Fig. S4.

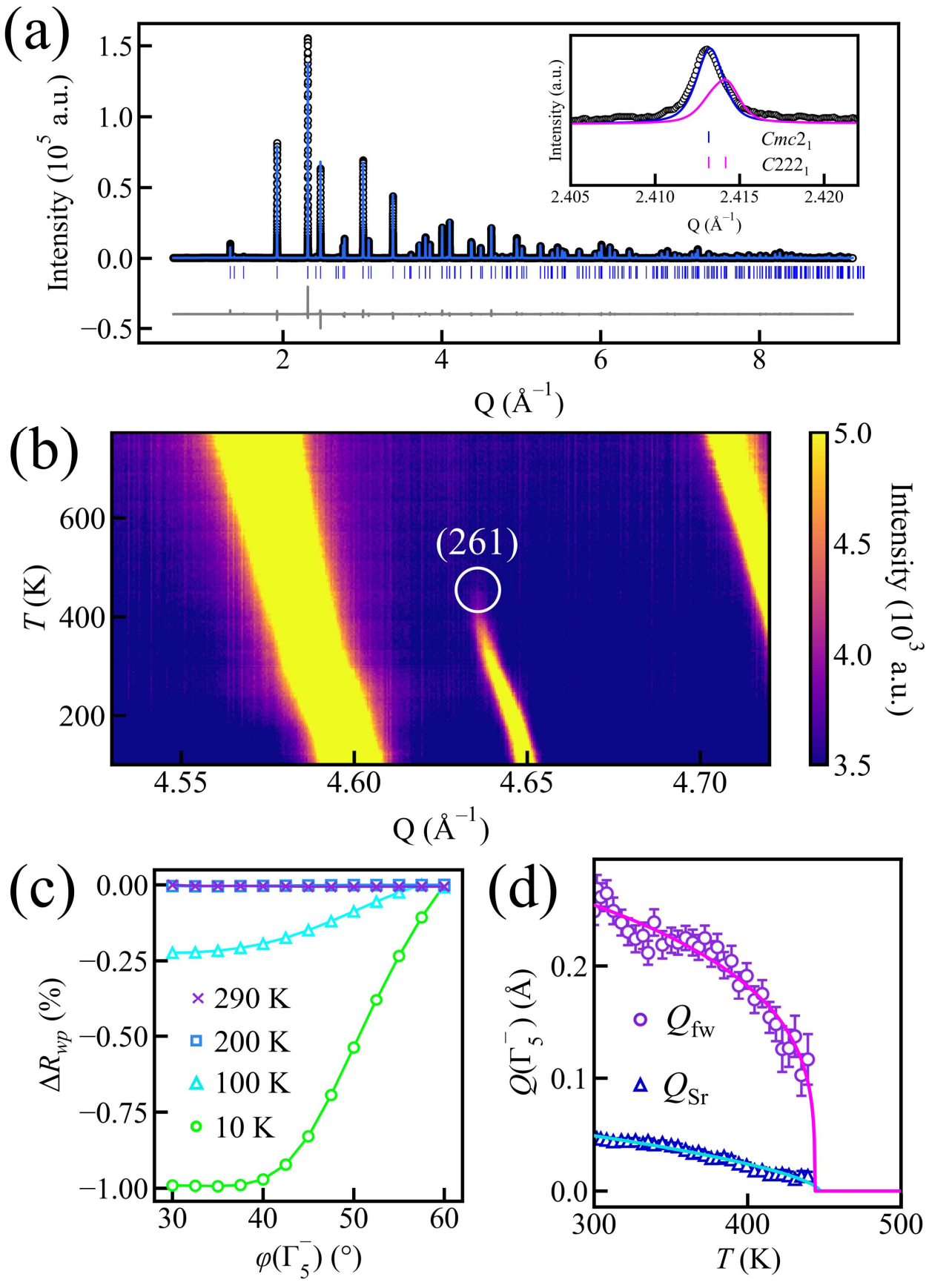

**Figure 2.** (a) Rietveld fit of the $Cmc2_1$ structural model against high-resolution S-XRD data collected at 10 K on the ID22 beamline ($\lambda = 0.3544921(1)$ Å). The inset compares the fits of the $C222_1$ (bottom ticks) and $Cmc2_1$ (top ticks) models for the (131) superstructure reflection (indexed

with respect to the $Cmc2_1$ structure). (b) Heat map of variable-temperature S-XRD data collected on the I11 beamline ($\lambda = 0.8287110(1)$ Å). The disappearance of the (261) superstructure reflection has been circled. (c) Significance tests based on structural models refined against ID22 data using fixed increments of the $\Gamma_5^-$ tilt axis ($\varphi(\Gamma_5^-)$) below $T_s$. $\varphi = 30°$ and $60°$ correspond to $Cmc2_1$ and $C222_1$ symmetries, respectively, while intermediate values correspond to $P2_1$ symmetry. The differences in $R_{wp}$ ($\Delta R_{wp}$) have been normalized with respect to the maximum $R_{wp}$ obtained. (d) The refined $\Gamma_5^-$ mode amplitudes decomposed in terms of the framework contribution ($Q_{fw}$) and the Sr displacements ($Q_{Sr}$). Solid lines depict fits against a power law expression $Q = A(T_s - T)^{\beta}$.

Our structural reinvestigation establishes that the $\Gamma_5^-$ tilts induce a spontaneous polarization in the 4H polytype. This reconciles recent reports of anomalous ferroelectric behavior in the 4H-$SrMnO_3$ system,[54–56] which were previously attributed to short-range compositional ordering or local structural fluctuations induced by chemical dopants (rather than the intrinsic, structurally driven mechanism we elucidate here). This interpretation is supported by our DFT calculations, which reveal that the lowest-frequency mode at the zone center corresponds to the $\Gamma_5^-$ RUM and it alone is sufficient to favor the $Cmc2_1$ phase, even in the absence of any intrinsic $\Gamma_2^-$ instability (**Figure 3**, Table S4). The $Cmc2_1$ phase is lower in energy than the $C222_1$ structure by ~0.092 meV/f.u., which is well above the energy resolution of our DFT calculations (~0.01 meV/f.u.). The relaxed $Cmc2_1$ structure is characterized by a small but measurable polarization of ~0.0245 μC cm$^{-2}$ (Table S4), comparable to that observed in magnetically induced ferroelectrics such as $TbMnO_3$.[57] However, in contrast to such systems where the polarization is contingent on the low-temperature magnetic order, the structural mechanism of the FE order in 4H-$SrMnO_3$ enables it to persist well above room temperature. This is due to the intrinsic stiffness associated with face-sharing polyhedra which favors higher phase transition temperatures,[58] so inversion-breaking

RUMs in other edge- or face-sharing framework structures will likely also exhibit high phase transition temperatures.

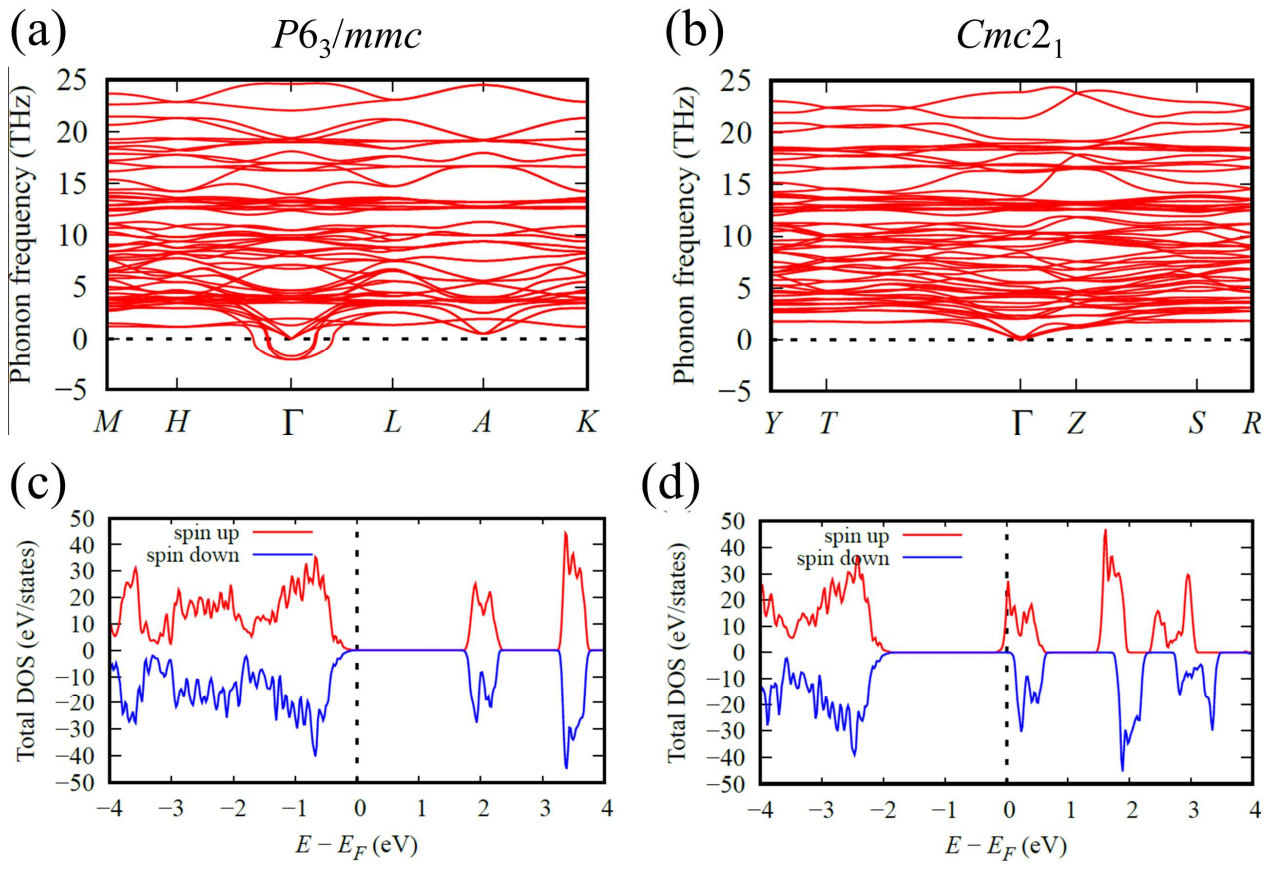

**Figure 3**. Phonon spectra calculated for (a) the high symmetry $P6_3/mmc$ structure and (b) the ground-state $Cmc2_1$ structure. (c), (d) Total density of states (TDOS) computed for the ground state $Cmc2_1$ structure without and with electron doping, respectively. Pristine 4H-$SrMnO_3$ is insulating with an electronic band gap of ~1.73 eV, while a small amount of electron doping (0.1 electrons per Mn) drives the system into a weakly metallic state.

While our results demonstrate that distortions transforming as $\Gamma_5^-$ induce FE order in 4H-$SrMnO_3$, this irrep is spanned by a total of six degrees of freedom (Table S2). Because the tilting of the $Mn_2O_9$ bioctahedra involves only one degree of freedom, it is necessary to verify that the observed FE distortion is driven primarily by the RUM itself, rather than by the other degrees of freedom transforming under the same irrep. Such non-RUM degrees of freedom may be characterized by a nontrivial compositional dependence that would frustrate efforts to systematically induce and tune the FE distortion. To this end, we decomposed the refined $\Gamma_5^-$ distortion in terms of a linear combination of symmetry-adapted displacement modes and compared them against the predicted displacement pattern of the $\Gamma_5^-$ RUM determined by the ISOTILT webtool (Supplementary Note 4). All six symmetry-adapted $\Gamma_5^-$ displacive modes were allowed to refine freely in our original refinements so as to avoid bias towards the expected RUM displacements. Despite this, we find that ~99% of the total $\Gamma_5^-$ distortion can be described by the renormalized RUM displacements alone (Table S8). The residual distortion predominantly stems from minor $Sr^{2+}$ displacements that do not form part of the octahedral framework, thus this represents the principal non-RUM degree of freedom transforming under the $\Gamma_5^-$ irrep which is active in the $Cmc2_1$ structure. Such *A*-site displacements commonly occur in tilted 3C perovskites and can be generally rationalized as an improper response to the primary framework distortion [7], so we conclude these Sr displacements have a similar physical origin in the 4H polytype as well.

To quantify the distinction between the framework and the *A*-site displacements, we decomposed the total mode amplitude obtained from symmetry-mode Rietveld refinements of variable-temperature I11 data (Supplementary Note 5) into the contributions from the $Mn_2O_9$ framework ($Q_{\text{fw}}$) and the Sr sites ($Q_{\text{Sr}}$) (Fig. 2d). By fitting $Q_{\text{fw}}$ and $Q_{\text{Sr}}$ separately to critical expressions of the form $Q \sim (T_{\text{S}} - T)^{\beta}$, we obtained estimates of $T_{\text{S}}$ and the critical exponent ($\beta$). The extracted transition temperatures are consistent within error ($T_{\text{S}}$ = 444(5) K for $Q_{\text{fw}}$ and 447(7) K for $Q_{\text{Sr}}$), but the critical exponents differ significantly. The framework component yields $\beta = 0.28(4)$, which is characteristic of a tricritical phase transition and consistent with the octahedral tilting transitions in 3C perovskites such as $SrZrO_3$.[59] In contrast, the Sr contribution is characterized by $\beta = 0.63(7)$, exceeding the typical range expected for an order parameter ($\beta \leq 0.5$). This behavior confirms that the Sr displacements effectively play an improper role in the phase transition and do not drive the observed FE order. Together, these results demonstrate that the stabilization of the $\Gamma_5^-$ RUM is responsible for the spontaneous polarization in this system.

Having established the role of the $\Gamma_5^-$ RUM in inducing the polar distortion, we now show that this same distortion can induce a wFM moment within the predominantly AFM spin texture. Previous studies reported that 4H-$SrMnO_3$ adopts an A-type AFM structure below $T_N \approx 280$ K, with the $Mn^{4+}$ spins aligned within the *ab* plane.[52] Our NPD measurements on GEM at ISIS (Fig. 4a) confirm this ordering, which transforms as the $m\Gamma_6^-$ irrep of the paramagnetic $P6_3/mmc$ structure (Fig. 1d) and is in good agreement with our first-principles calculations (Supplementary Note 3). Unlike the canonical ME perovskite $BiFeO_3$,[60] we detect no magnetic diffuse scattering or satellite reflections down to 1.5 K to suggest that a non-collinear magnetic structure manifests. However, DC magnetic susceptibility measurements (Fig. 4b) reveal a clear divergence between zero-field-cooled (ZFC) and field-cooled (FC) responses below $T_N$, indicative of a wFM component that we are insensitive to in our NPD data. Hysteresis loops recorded in the range 200–300 K show that, although there is a very weak FM moment within the sample above $T_N$ (Fig. S15a), there is a clear enhancement in the remanent magnetization ($M_r$) below $T_N$ (Fig. S15b) so that the AFM transition is accompanied by its own wFM component.

Symmetry analysis (Supplementary Note 6) shows that the observed wFM moment must arise from coupling between the $m\Gamma_6^-$ AFM order and either the $\Gamma_5^-$ tilt or $\Gamma_2^-$ polar modes as the $m\Gamma_6^-$ AFM order by itself is forbidden by symmetry to produce a wFM moment. To verify this, DC susceptibility measurements on a freshly synthesized sample of the untilted 4H-$BaMnO_3$ composition (Supplementary Note 7) showed no evidence of an equivalent wFM component below $T_N$, despite it exhibiting the same A-type AFM order in its ground state.[51] Since $T_N$ in 4H-$SrMnO_3$ is preceded by a regime of short-range AFM ordering,[61,62] coupling via the tilts and/or the polar distortion likely also explains the residual $M_r$ we detect above $T_N$ (Fig. S15b).

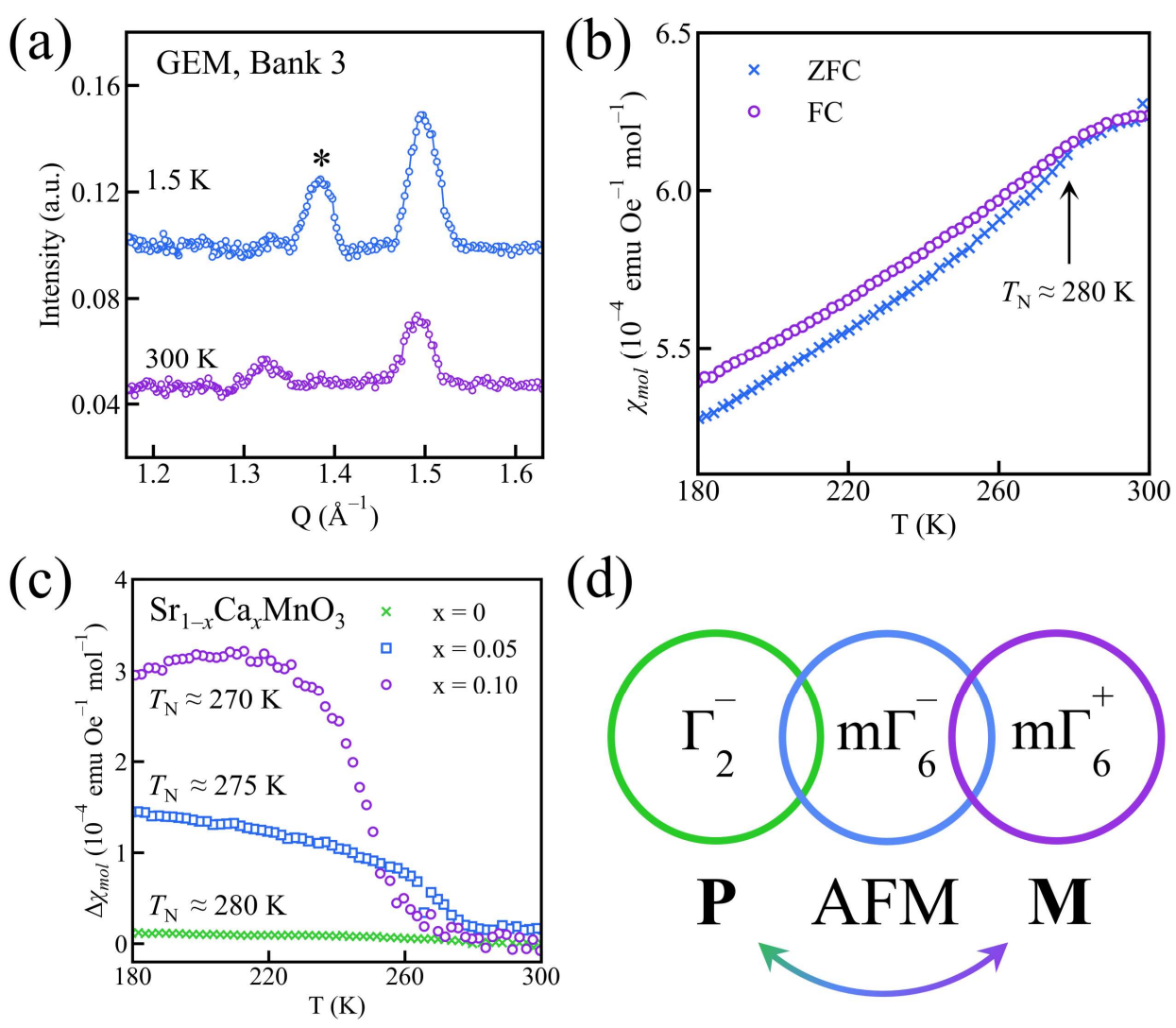

**Figure 4.** Magnetic properties of $SrMnO_3$. (a) NPD data collected on Bank 3 of GEM, highlighting the emergence of the $(002)_{mag}$ reflection at 1.5 K (*). (b) Molar DC susceptibility ($\chi_{mol}$) of 4H-$SrMnO_3$, showing the ZFC-FC divergence at $T_N$. (c) ZFC-FC divergence in $\chi_{mol}$ ($\Delta\chi_{mol}$) for the series 4H-$Sr_{1-x}Ca_xMnO_3$, showing the enhancement of the wFM moment with increasing $Ca^{2+}$ substitution. (d) ME coupling in 4H-$SrMnO_3$, showing how polarization (**P**) and magnetization (**M**) form a trilinear coupling via the AFM mode.

Symmetry constraints uniquely require the induced wFM moment to transform as the $m\Gamma_6^+$ irrep (Fig. 1e), which is the only mode compatible with the site symmetries of the 4H structure to yield a net ferromagnetic moment. Our non-collinear spin-polarized DFT calculations with SOC confirm that the magnetic easy axis lies along the crystallographic *x*-direction in the $Cmc2_1$ phase. This permits a wFM $m\Gamma_6^+$ mode (Fig. 1e) yielding Mn moments parallel to those produced by the primary $m\Gamma_6^-$ AFM order, producing a weak ferrimagnetic (wFiM) ground state described by the Shubnikov group $Cmc'2_1'$ (Fig. 5a). However, although the wFiM state is allowed by symmetry, the insulating nature of 4H-$SrMnO_3$ (**Figure 3**c, Fig. S17a) and the absence of charge ordering do not favor a sizeable imbalance between the antiparallel Mn Moments, and for this reason the resulting wFM moment obtained from DFT remains very small (~0.001 $\mu_B$). To verify this, we performed charged DFT calculations by adding 0.1 electrons per Mn atom – this drives the system to become weakly metallic, as shown in **Figure 3**d. Interestingly, even this small level of electron doping induces a finite imbalance between the magnetic moments of the two inequivalent Mn sites within a $Mn_2O_9$ bioctahedral dimer, leading to a non-zero $m\Gamma_6^+$ wFM mode with an amplitude of ~0.219 $\mu_B$. These results suggest that carrier doping, possibly either through introducing oxygen vacancies or aliovalent substitutions on the *A* and/or *B* sites, could enhance the symmetry-allowed wFM instability and may provide a promising route toward stabilizing the wFiM state. Our preliminary attempts to prepare reduced $SrMnO_{3-x}$ samples confirmed that electron doping leads to a modest enhancement in the wFM moment (Fig. S16), demonstrating the validity of this approach.

Although carrier doping can enhance the wFM moment, this will drive the system metallic so any enhancement will come at the expense of practical FE/ME switching. For this reason, we sought alternative chemical modifications that could enhance the wFM moment without sacrificing the insulating ground state. Because the wFM order is coupled to the $\Gamma_5^-$ tilts, increasing the tilt magnitude provides a potential route to strengthen the wFM response. To demonstrate this, we found that the partial substitution of 5% and 10% $Ca^{2+}$ ($r$ = 1.34 Å) for $Sr^{2+}$ ($r$ = 1.44 Å) systematically enhances the wFM component by at least one order of magnitude (Fig. 4c). This can be rationalized by the smaller ionic radius of $Ca^{2+}$, which increases *A*-site underbonding [50] and promotes larger tilting of the $Mn_2O_9$ bioctahedra [46]. Importantly, these modest $Ca^{2+}$ substitutions avoid increasing the electronic conductivity while preserving the appealingly high Néel temperature in this system, with $T_N \approx 270$ K for $Sr_{0.9}Ca_{0.1}MnO_3$ (Fig. 4c). Clearly, the interdimer Mn–O–Mn superexchange interactions are only weakly perturbed by the isovalent substitution of $Ca^{2+}$ for $Sr^{2+}$, highlighting the strong tuneability of the wFM response through chemical modifications. The improper nature of the induced polarization should also allow compressive strain to enhance both the tilt amplitudes and the polarization/wFM responses, as has been achieved in $TbMnO_3$.[63] This is in contrast to what is typically observed in canonical hybrid-improper ferroelectrics such as $Ca_3Ti_2O_7$, where a weak proper component due to the $Ti^{4+}$ cations suppresses the polarization upon the application of hydrostatic pressure.[64]

The semiconducting nature of our samples precludes direct observation of ME switching (Supplementary Note 8). However, the symmetry basis of our design strategy allows us to identify the precise nature of the ME coupling (Fig. 4d) and propose a plausible ME switching pathway by analyzing the Landau free energy expansion with respect to the high-symmetry paramagnetic phase (Supplementary Note 9). The lowest-order coupling between the $\Gamma_2^-$ polar mode and the AFM/wFM components takes the form:

$$F \sim PM_1M_2 \sin\Delta, \qquad \Delta = \theta_1 - \theta_2 \tag{1}$$

where $M_1$ and $M_2$ are the magnitudes of the symmetrized $m\Gamma_6^-$ (AFM) and $m\Gamma_6^+$ (wFM) order parameter components, respectively, and $\theta_1$ and $\theta_2$ are their corresponding phase angles. This closely resembles the form of the ME coupling found in magnetically driven ferroelectrics such as $Ni_3V_2O_8$,[65] but we emphasize the polarization in 4H-$SrMnO_3$ is structurally induced (rather than magnetically) via a separate coupling term of the form $F \sim P\rho^3\sin3\varphi$, where $\rho$ and $\varphi$ are the magnitude and phase of the $\Gamma_5^-$ order parameter (Supplementary Note 9). For the $Cmc'2_1'$ ground state determined from our DFT calculations, the relative phase shift between the magnetic order parameter directions is $\Delta = \pm 90°$. Physically, this corresponds to an exact parallel alignment between the AFM and wFM moments, resulting in a net wFiM state (Fig. 5a). The linear dependence on $P$ in Eq. (1) permits the system to adopt either a positive or negative polarization to minimize the free energy, so to maintain invariance overall the transformation $P \rightarrow -P$ necessarily switches Δ by ±180° (Fig. 5b). Since the switching barrier associated with reorienting the AFM component is likely to be much larger compared to the wFM component, this is most plausibly achieved by reorienting $\theta_2$ by ±180°, resulting in a redistribution of the magnetic imbalance within the $Mn_2O_9$ bioctahedral dimers. Although the real switching pathway is likely to be far more complex than this (see Supplementary Note 9), this coupling nonetheless establishes a feasible basis for the switching of net magnetization under an applied electrical field.

Physically, the crystallographic inequivalence of the $Mn^{4+}$ sites within the bioctahedral dimers is a direct consequence of the improper $\Gamma_2^-$ polar distortion, which splits the single $Mn^{4+}$ site in the parent $P6_3/mmc$ structure into two. This is what permits the imbalance between the $Mn^{4+}$ moments, meaning the induced wFM moment relies on the spontaneous polarization induced within the crystal structure. We confirmed this by performing additional charged DFT calculations for the $C222_1$ phase – although the system becomes weakly metallic upon electron doping, we found no imbalance between the $Mn^{4+}$ moments to suggest that any wFiM state is stabilized. This suggests the spin splitting in the doped $Cmc2_1$ phase is indeed symmetry-driven and does not arise from d-electron correlations. Combined, this demonstrates the precise mechanistic link between the $\Gamma_5^-$ tilts, the $\Gamma_2^-$ polar distortion, and the $m\Gamma_6^+$ wFM moment (Fig. 1f).

From a design perspective, the key advantage of 4H-$SrMnO_3$ lies in the simplicity of its ME coupling scheme (Fig. 4d). All relevant order parameters transform as zone-centered (Γ-point) irreps of the parent structure, ensuring conservation of crystal momentum and enabling direct coupling between the structural, polar, and magnetic modes. This contrasts with established hexagonal multiferroics such as $YMnO_3$,[66,67] where the coupling involves a zone-boundary $K_3$ distortion [68] and $mK_3$ AFM order, formally precluding a net magnetization within the bulk.[69] Similarly, in conventional 3C perovskites, the high symmetry of the $Pm\bar{3}m$ aristotype structure requires multiple zone-boundary modes to be condensed simultaneously to achieve comparable coupling.[13] In contrast, the Γ-point coupling scheme realized here requires only a minimal set of symmetry-breaking distortions – namely, the stabilization of the $\Gamma_5^-$ tilts (to induce the $\Gamma_2^-$ polar mode on an improper basis), and A-type AFM order (to allow for a wFM component). Taken together, these results reveal an unusually simple and complete realization of ME coupling within a single symmetry-based framework. To our knowledge, 4H-$SrMnO_3$ represents the first experimentally realized system that satisfies such a simple yet robust ME coupling mechanism.[70,71] Similar behavior has been predicted recently in oxygen-deficient Ruddlesden–Popper phases of the form $A_4B_3O_9$,[72] although experimental realization of this remains outstanding. Nevertheless, 4H-$SrMnO_3$ provides the essential crystallochemical ingredients required for robust ME coupling which, coupled with its appealingly high structural and magnetic ordering temperatures, presents promising opportunities for the realization of room-temperature ME switching.

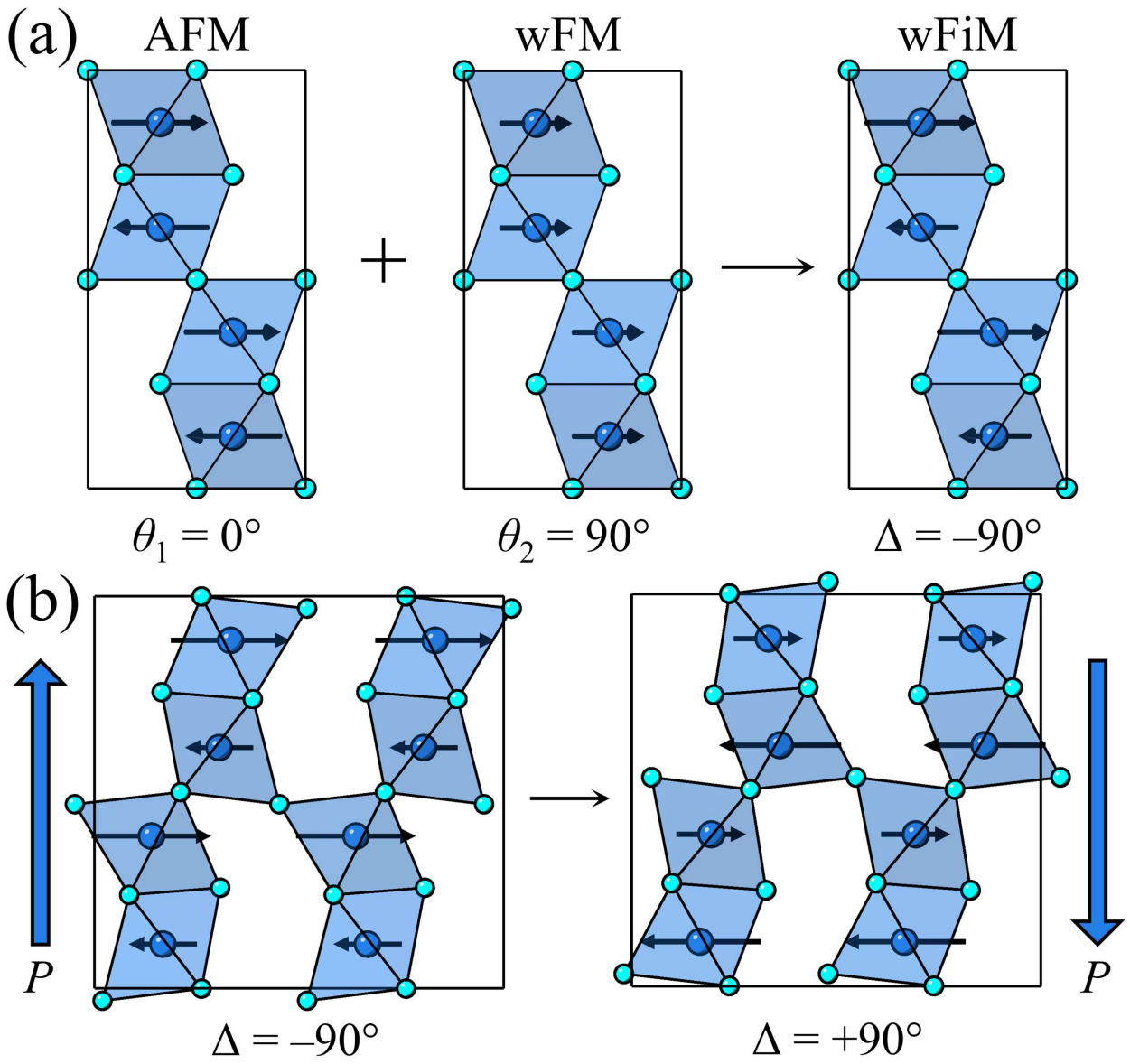

Figure 5. (a) Symmetry-driven generation of a weak ferrimagnetic (wFiM) state via the combination of parallel AFM and wFM moments in the *Cmc*'$2_1$' magnetic structure. (b) A plausible ME switching pathway, showing the redistribution in magnetic moments across the crystallographically inequivalent $Mn^{4+}$ sites. The blue arrows clarify the directions of the net polarization within the structures. The total amplitude of the structural distortion in (b) has been amplified by a factor of two for visual clarity.

**CONCLUSIONS**. Our high-resolution diffraction, symmetry-mode analysis, and DFT calculations collectively establish that a single $\Gamma_5^-$ RUM simultaneously induces a spontaneous polarization and a wFM moment in the 4H polytype. To the best of our knowledge, this is the first example of a single RUM in a crystalline framework capable of breaking global inversion symmetry without the influence of extraneous symmetry-breaking distortions. Similar inversion-breaking RUMs could arise in other framework structures deviating from the conventional perovskite architecture, thus presenting a promising new strategy to design coupled FE and wFM orders near room temperature. The present 4H-$A$MnO$_3$ system shows that this can be achieved with minimal symmetry-breaking ingredients, and additional modifications to the *A*/*B*-site size mismatch and/or manipulations of the defect chemistry present further opportunities to enhance the functional properties while preserving the appealingly high ordering temperatures. Although our present work has targeted a ternary ceramic oxide, one of the most appealing facets of our design approach is that it may be readily generalized to other material chemistries. Both halide and hybrid perovskites are of particular interest in this respect due to their propensity to form polytype structures, and the design strategy we outline here can likely be extended to tune the optoelectronic properties which are of timely interest in these materials. Our work now stimulates the discovery of similar inversion-breaking RUMs in other framework structures that can potentially imbue them with newfound ferroic and/or ME functionalities. Ultimately, this paves new pathways for the rational design of ferroelectric and magnetoelectric materials across more complex chemical architectures than has been possible previously.

ASSOCIATED CONTENT

**Supporting Information**. The following files are available free of charge.

Further details on the symmetry analysis, Rietveld refinements, DFT calculations, RUM analysis, magnetic/physical property characterization, and Landau analysis (PDF)

$Cmc2_1$ structural models refined against synchrotron XRD data collected at 10 K and 290 K (.cif)

Landau free energy expansion up to sixth order for the $\Gamma_5^-$, $\Gamma_2^-$, $m\Gamma_6^-$, and $m\Gamma_6^+$ order parameters (.txt).

AUTHOR INFORMATION

**Corresponding Authors**

Struan Simpson; Email: struan.simpson@warwick.ac.uk

Mark Senn; Email: m.senn@warwick.ac.uk

**Author Contributions**

The manuscript was written through contributions of all authors. All authors have given approval to the final version of the manuscript.

**ACKNOWLEDGMENTS**

S.S. and M.S.S. acknowledge funding from the Royal Society (UF160265 and URF\R\231012). Preliminary sample characterization was supported by the Warwick X-ray Research and Technology Platform. We acknowledge the European Synchrotron Radiation Facility (ESRF) for the provision of beamtime allocated on the ID22 beamline under proposal number MA-6810 (doi:10.15151/ESRF-ES-2220630641), Diamond Light Source for beamtime allocated on the I11 beamline under the Oxford-Warwick Solid-State Chemistry BAG (CY32893 and CY39378), and the ISIS Neutron and Muon Source for beamtime allocated on the GEM diffractometer under proposal number RB2510392 (doi:10.5286/ISIS.E.RB2510392). We thank A. N. Fitch and C. Dejoie for support in using ID22 at the ESRF, as well as S. P. Thompson, E. Connolly and S. Day for support in using I11 at Diamond Light Source. U.D. acknowledges useful discussions with Jorge Íñiguez-González and financial support from the Luxembourg National Research Fund (FNR) through grant C23/MS/17909853/BUBBLACED. This work made use of the high-performance computing (HPC) facilities of the University of Luxembourg [73] (https://hpc.uni.lu).

For Table of Contents Only

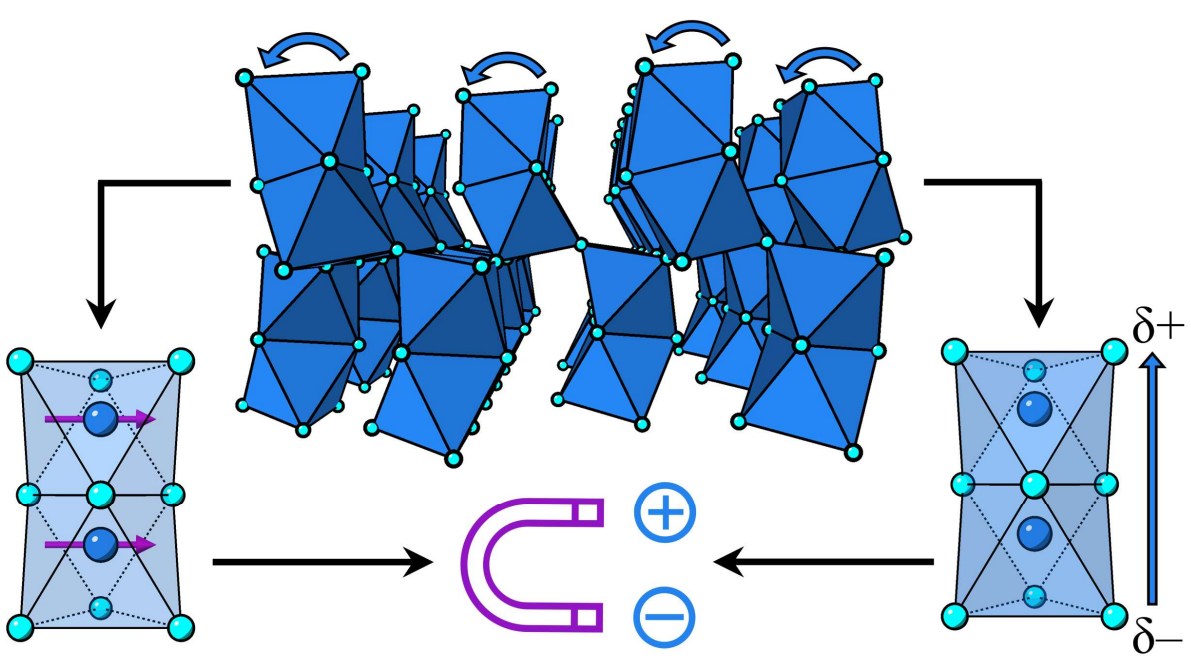

Supplementary Information for: "Near-room-temperature magnetoelectric coupling engineered through inversion-breaking tilts in a bulk perovskite polytype"

## Supplementary Note 1. Relationship between the $\Gamma_5^-$ tilt axis and the structural symmetry.

Figure S6 depicts the relationship between the order parameter space of the $\Gamma_5^-$ tilt mode and the symmetry of the crystal structure. The full basis spanned by the $\Gamma_5^-$ (a,b) irrep encompasses $C222_1$, $Cmc2_1$, and $P2_1$ symmetries, which correspond to representative order parameter directions (OPDs) of (a,0), (a,$\sqrt{3}$a/3), and (a,b), respectively. The high-symmetry OPDs can be mapped in terms of two sets of sixfold structural domains possessing either $C222_1$ or $Cmc2_1$ symmetry. By writing the OPD in terms of a phase angle $\varphi = \tan^{-1}(b/a)$, these correspond to $\varphi = n\pi/3$ and $(2n+1)\pi/6$, respectively, where $n$ is an integer, hence symmetry-equivalent structural domains are related to each other by $\Delta\varphi = \pm\pi/3$. Physically, the differences between these structures can be understood on the basis of the orientation of the tilt axis within the hexagonal basal plane (Figure S6b and S1c). Due to the face-sharing geometry between pairs of octahedra, the octahedral *B*-site atoms necessarily displace perpendicularly to the tilt axis as part of any proposed rotation (see Fig. 1c in the main text). The displacements are antiparallel across hexagonal close-packed $AX_3$ layers (i.e., across the face-sharing octahedra), while they are parallel across cubic close-packed $AX_3$ layers (i.e., across the corner-sharing octahedra).

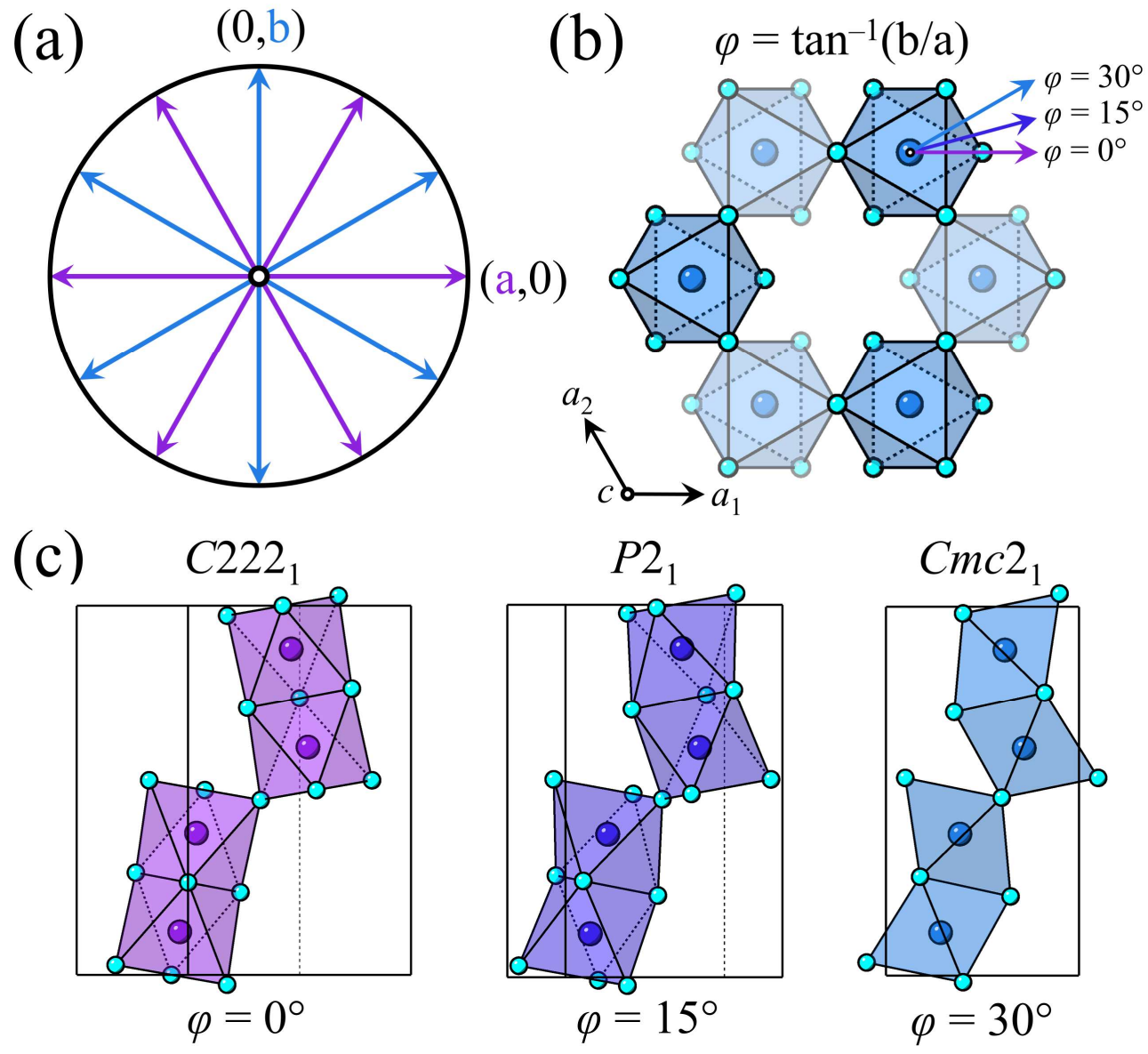

Figure S6. (a) The full order parameter space spanned by the $\Gamma_5^-$ (a,b) irrep. Symmetry-related domains are depicted using arrows of the same color, with purple and blue representing $C222_1$ and $Cmc2_1$ domains, respectively. (b) Partial map of the $\Gamma_5^-$ order parameter space projected onto the crystal structure of the 4H polytype viewed down the *c* axis. Three representative tilt axes have been depicted in terms of $\varphi = \tan^{-1}(b/a)$. The unit cell axes are labelled with respect to the hexagonal parent structure. (c) Views of the crystal structure along the three tilt axes shown in (b). The parent hexagonal cell has been superimposed on each, and the *A*-site atoms have been omitted for visual clarity.

Supplementary Note 2. Determining the ground-state crystal structure of 4H-$SrMnO_3$.

Previous diffraction experiments concluded that the ground-state structure of 4H-$SrMnO_3$ was best described by $C222_1$ symmetry [1]. This prior space group assignment was made on the basis of the appearance of $hh(2\bar{h})l$: $l = 2n + 1$ reflections in selected-area electron diffraction patterns, combined with systematic tests of different structural models against neutron and synchrotron diffraction data based on higher-order isotropy subgroups of the parent $P6_3/mmc$ space group. Based on the $C222_1$ assignment that was made, it was concluded that the low-symmetry structure could be described in terms of its transformation as the $\Gamma_5^-$ (a,0) irrep of the $P6_3/mmc$ structure. Our own S-XRD experiments on the ID22 beamline showed that this $C222_1$ structural model gave a broadly reasonable description of the observed data (Figure S7), but it fit poorly against the superstructure reflections (Figure S8). We thus sought to re-examine the structural assignment below $T_s$ to deduce whether an alternative choice of space group was more appropriate.

The $C222_1$ and $Cmc2_1$ space groups are both $C$-centered, hence they are distinguished by the general reflection condition $hkl$: $h + k = 2n$ (where $n$ is an integer). However, the $Cmc2_1$ space group is distinguished from $C222_1$ by the additional condition $h0l$: $h,l = 2n$, hence an additional set of superstructure reflections is predicted for the $C222_1$ structure compared to the $Cmc2_1$ structure. This provides a systematic basis from which to deduce the correct space group assignment. In general, this leads to the $C222_1$ and $P2_1$ models predicting a splitting of the superstructure reflections, while no such splitting is predicted for the $Cmc2_1$ model (Figure S8). We do not find convincing evidence of such splitting, nor any anomalous broadening with respect to the parent Bragg reflections, within the exceptionally high resolution of our S-XRD experiments ($\Delta d/d \approx 1 \times 10^{-4}$). To verify this, we tested all possible space group assignments spanned by the $\Gamma_5^-$ (a,b) basis (Figure S6) against our S-XRD data.

Our structural models were constructed as follows. The instrumental contribution to the peak shape was modelled by a simple convolution of fixed Gaussian and Lorentzian contributions, each obtained by refinements against data collected on a Si standard. 21 background terms were refined, including 12 terms of a shifted Chebyschev polynomial and 9 additional parameters describing three Gaussian peaks to account for background contributions from the cryostat. Both a zero-point error and a scale factor were refined in addition to the lattice parameters (3 for the $C222_1$ and $Cmc2_1$ models and 4 for the $P2_1$ model). The internal atomic coordinates were refined freely for each model, with 10, 18, and 30 independent degrees of freedom for the $C222_1$, $Cmc2_1$, and $P2_1$ structures, respectively. Isotropic displacement parameters were refined for each atomic site and constrained to be equal for each element; the only exception was the Sr(1) site, where anisotropic displacement parameters consistently improved the quality of fit (~2% reduction in $R_{wp}$) for each model. Overall, this resulted in 7 additional refined parameters for the $C222_1$ and $Cmc2_1$ models and 9 for the $P2_1$ model. Finally, we also included 2 crystallite size broadening terms as well as Stephens anisotropic broadening terms (7 parameters for $C222_1$ and $Cmc2_1$; 10 for $P2_1$) to model sample-dependent contributions to the peak shapes. This resulted in a total of 52, 60, and 78 refined parameters for the $C222_1$, $Cmc2_1$, and $P2_1$ models, respectively. These were then fitted against the data collected at 10 K, 100 K, 200 K, and 290 K, thus spanning a range of temperatures below $T_s$ to ensure sensitivity to the correct space group assignment.

The fitting statistics obtained from each model are summarized in Table S1. At 290 K and 200 K, we find that little distinguishes the three possible choices of structural model, potentially explaining prior ambiguity regarding the true space assignment below $T_s$. However, at 100 K and 10 K, we find that the differences between the three structural models become more pronounced. This directly reflects the enhanced ferroelastic (i.e., orthorhombic) strain at low temperatures, hence powder diffraction provides us with sufficient sensitivity to distinguish between the different structural models. Based on the combination of fit quality and the number of fitting parameters, our Rietveld fits show that the $Cmc2_1$ model provides the best description of the low-temperature data. This is further supported by its ability to describe the low-temperature superstructure reflections more accurately than the canonical $C222_1$ model (Figure S8). The $P2_1$ model is distinguished from the $Cmc2_1$ model by an additional 18 refined parameters, so the modest improvement for this model can be attributed to overparameterization.

Table S1. Summary of fitting statistics for Rietveld-fits performed against synchrotron XRD (S-XRD) data at select temperatures below the structural phase transition. The $C222_1$, $Cmc2_1$, and $P2_1$ models had 52, 60, and 78 refined parameters, respectively.

| $T$ (K) | Model | S-XRD $R_{wp}$ (%) |
|---|---|---|
| 290 | $C222_1$ | 7.29 |
| | $Cmc2_1$ | 7.27 |
| | $P2_1$ | 7.11 |
| 200 | $C222_1$ | 7.73 |
| | $Cmc2_1$ | 7.65 |
| | $P2_1$ | 7.29 |
| 100 | $C222_1$ | 8.38 |
| | $Cmc2_1$ | 7.79 |
| | $P2_1$ | 7.38 |
| 10 | $C222_1$ | 9.28 |
| | $Cmc2_1$ | 8.15 |
| | $P2_1$ | 7.40 |

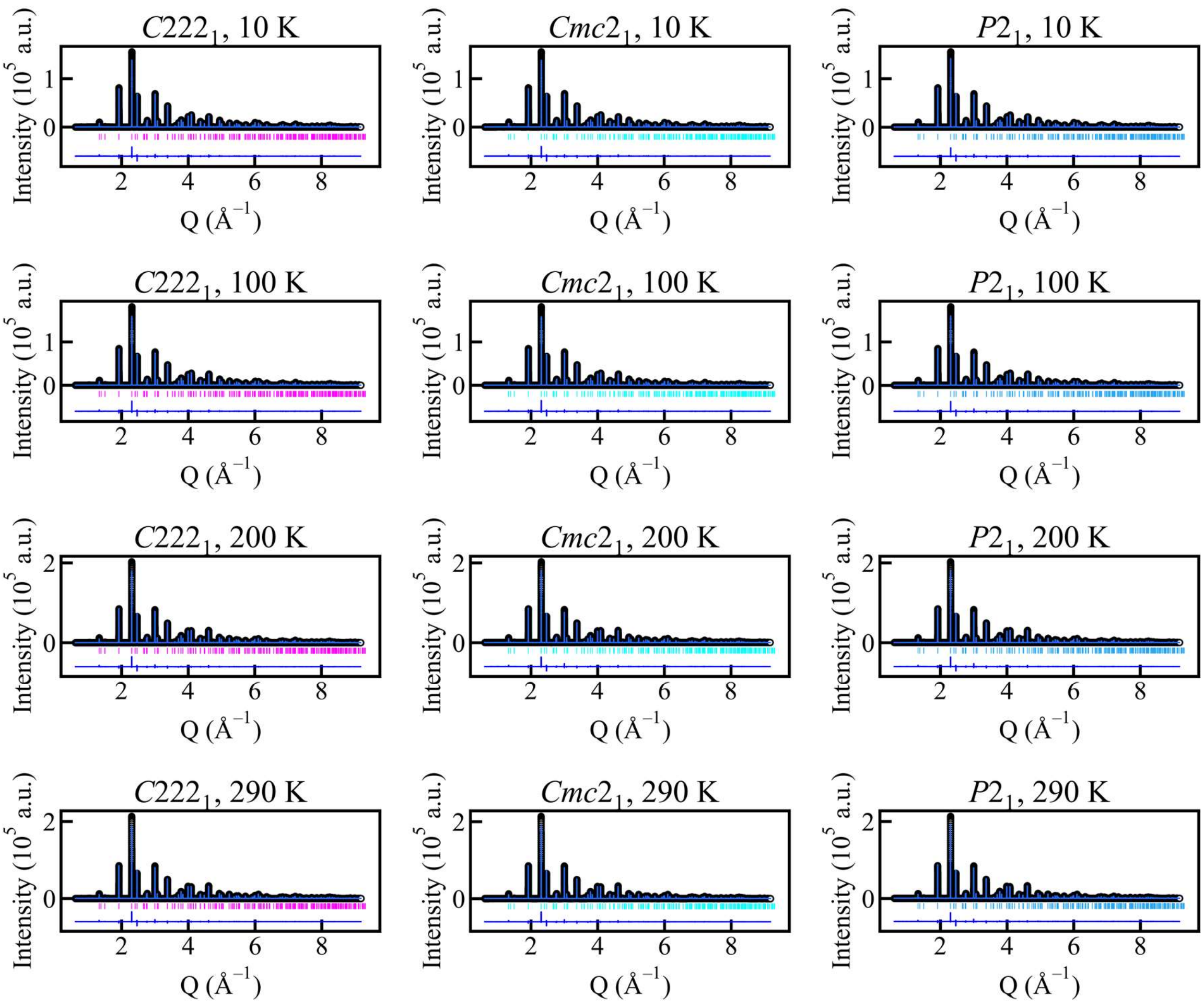

Figure S7. Rietveld fits of the $C222_1$, $Cmc2_1$, and $P2_1$ structural models against S-XRD data collected on ID22 at select temperatures below $T_s$. Pink, cyan, and blue tick marks represent the positions of the expected Bragg reflections for the $C222_1$, $Cmc2_1$, and $P2_1$ models, respectively.

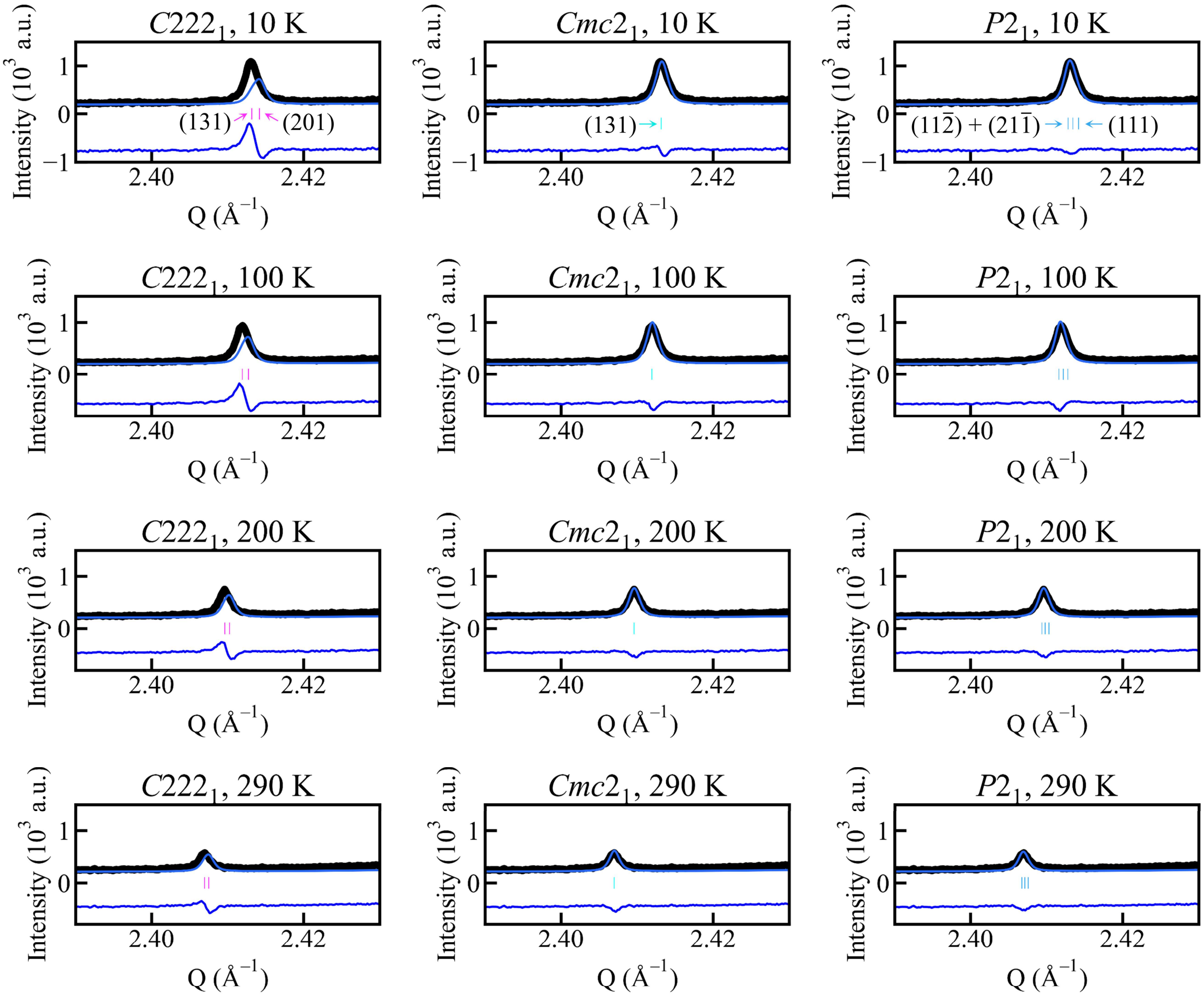

Figure S8. Expanded view of the Rietveld fits against the ID22 data, highlighting the poor fit of the $C222_1$ structural model and the good fits of the $Cmc2_1$ and $P2_1$ models against the first observed superstructure reflection. The predicted reflections for each model have been highlighted and indexed with respect to their corresponding space groups in the 10 K pattern. The $C222_1$ and $P2_1$ structures both predict a subtle splitting of the superstructure reflections, while the $Cmc2_1$ model predicts only a single superstructure reflection. At 10 K, the ultra-high resolution of our S-XRD data ($\Delta d/d \approx 1 \times 10^{-4}$) is sufficient to allow us to distinguish any splitting or broadening of the superstructure reflections, but as shown by the fits against the $C222_1$ and $P2_1$ model, this is either not consistent with the observed pattern or unnecessary to model the observed superstructure reflections. Any departure from the orthorhombic metric symmetry must therefore comprise of a strain of less than $10^{-4}$.

Normally, the challenge of distinguishing the precise orientation of the tilt axis with respect to the crystal structure would be intractable by Rietveld refinement alone as the tilts predominantly involve displacements of the oxygen atoms (Table S2). This means the $C222_1$, $Cmc2_1$, and $P2_1$ models will be mostly distinguished by subtle intensity differences in the calculated patterns, and the peak intensities will be strongly correlated to the other refined parameters in our model. However, the enhanced sensitivity of S-XRD to precise peak positions allows us to exploit the improper ferroelastic strain transforming as the two-dimensional $\Gamma_5^+$ (c,d) irrep to infer the $\Gamma_5^-$ OPD; this is precisely why we make our structural assignment on the basis of our S-XRD data, rather than the NPD data (despite the greater sensitivity of the latter to the precise oxygen positions). The $C222_1$ and $Cmc2_1$ structures may be mapped to distinct $\Gamma_5^+$ OPDs ((a,$-\sqrt{3}$a) and (a,0), respectively). To encode this in our refinements, we parameterized the $\Gamma_5^+$ strain as an order parameter of the form ($\varepsilon$, $\theta$), where $\varepsilon = (c^2 + d^2)^{1/2}$ and $\theta = \tan^{-1}(d/c)$. Much like the $\Gamma_5^-$ tilt mode, this allows us to write the $\Gamma_5^+$ OPD in terms of a phase angle $\theta$ and construct a series of structural models which linearly interpolate between the $C222_1$ and $Cmc2_1$ OPDs in terms of $\theta$. A trivial mapping between the bases of the $\Gamma_5^+$ and $\Gamma_5^-$ irreps then allows us to rewrite $\theta$ in terms of the $\Gamma_5^-$ OPD (i.e., $\varphi$) as $\theta = 2\varphi - \pi/3$. Thus, although our models are strictly speaking parameterized with respect to the $\Gamma_5^+$ strain, they can be directly interpreted in terms of the $\Gamma_5^-$ OPD. Note that, due to the polycrystalline nature of our samples, we are insensitive to the absolute orientation of the $\Gamma_5^+$ OPD. Nevertheless, we retain sensitivity to its relative orientation so that we can still distinguish the average crystallographic symmetry across the bulk with this approach.

To perform our significance tests, our refinements were constructed as follows. The internal atomic degrees of freedom were fixed based on the displacive mode components obtained from a $Cmc2_1$ symmetry-mode refinement performed against the 10 K dataset. Since changes to these displacement modes only effect minor intensity differences to the calculated patterns, this constraint minimizes the number of refined parameters and helps to ensure any significant changes in fitting statistics can be attributed to the systematic variation in $\theta$ (hence, $\varphi$) across the different structural models. We also chose to perform our refinements using isotropic displacement parameters that were constrained based on their refined values at 10 K. We then refined a global scaling parameter for these displacement parameters for the higher-temperature datasets. The anisotropic Stephens terms were also replaced with simple isotropic strain broadening terms to minimize the total number of refined parameters in our model. Ultimately, the only freely refined parameters in our significance tests were the background terms, the zero-shift and scale parameters, the $\Gamma_1^+$ strain modes (to account for thermal expansion), and $\rho(\Gamma_5^+)$. Additionally, each structural model corresponding to different choices of $\varphi$ was refined on an individual basis (rather than a sequential basis) so as to avoid introducing bias into the refinements.

Fig. 2c shows the results of our significance tests at select temperatures below 290 K. At 10 K, we find that the refinement surface is characterized by a broad minimum around the $Cmc2_1$ OPD ($\varphi$ = 30°), while the $C222_1$ OPD ($\varphi$ = 60°) represents a clear maximum. This is also true for the data collected at 100 K, suggesting $Cmc2_1$ symmetry persists up to at least this temperature. The differences between the two high-symmetry OPDs are much smaller at higher temperatures. This is due to the reduction in magnitude of the ferroelastic strain ($\varepsilon$ at 200 K is ~20% of the 10 K value), so we are less sensitive to the precise space group assignment at higher temperatures. Nevertheless, the appearance of the superstructure reflections remains broadly unchanged up to 290 K (Figure S8), and we find no evidence of anomalous broadening compared to the parent Bragg reflections at higher temperatures to suggest there is any additional phase transition away from the polar structure within the very high resolution of our S-XRD experiments ($\Delta d/d \approx 1 \times 10^{-4}$). Therefore, our significance tests demonstrate that the $Cmc2_1$ model is the most appropriate structural assignment for the tilted structure of 4H-$SrMnO_3$.

The final fitting parameters obtained from our $Cmc2_1$ structural model are shown in Figure S9. The transformation matrix relating the parent lattice vectors of the $P6_3/mmc$ structure to the $Cmc2_1$ basis is:

$$\begin{pmatrix} 1 & 0 & 0 & 0 \\ 1 & 2 & 0 & 0 \\ 0 & 0 & 1 & 0 \end{pmatrix}.$$

We find no evidence of any volume discontinuity across the full temperature range spanned by our S-XRD experiments, substantiating that no additional phase transitions occur up to $T_s$. This is in contrast to the 6H polytype of $BaTiO_3$, where its $Cmc2_1 \rightarrow C222_1$ transition is marked by a clear ferroelastic anomaly as well as an exotic Goldstone regime which is readily distinguished by significant diffuse-like scattering between twin-related diffraction peaks [2]. The amplitudes of the refined $\Gamma_5^-$ components at 10 K are provided in Table S2, and the Mn–O bond lengths are provided for reference in Table S3.

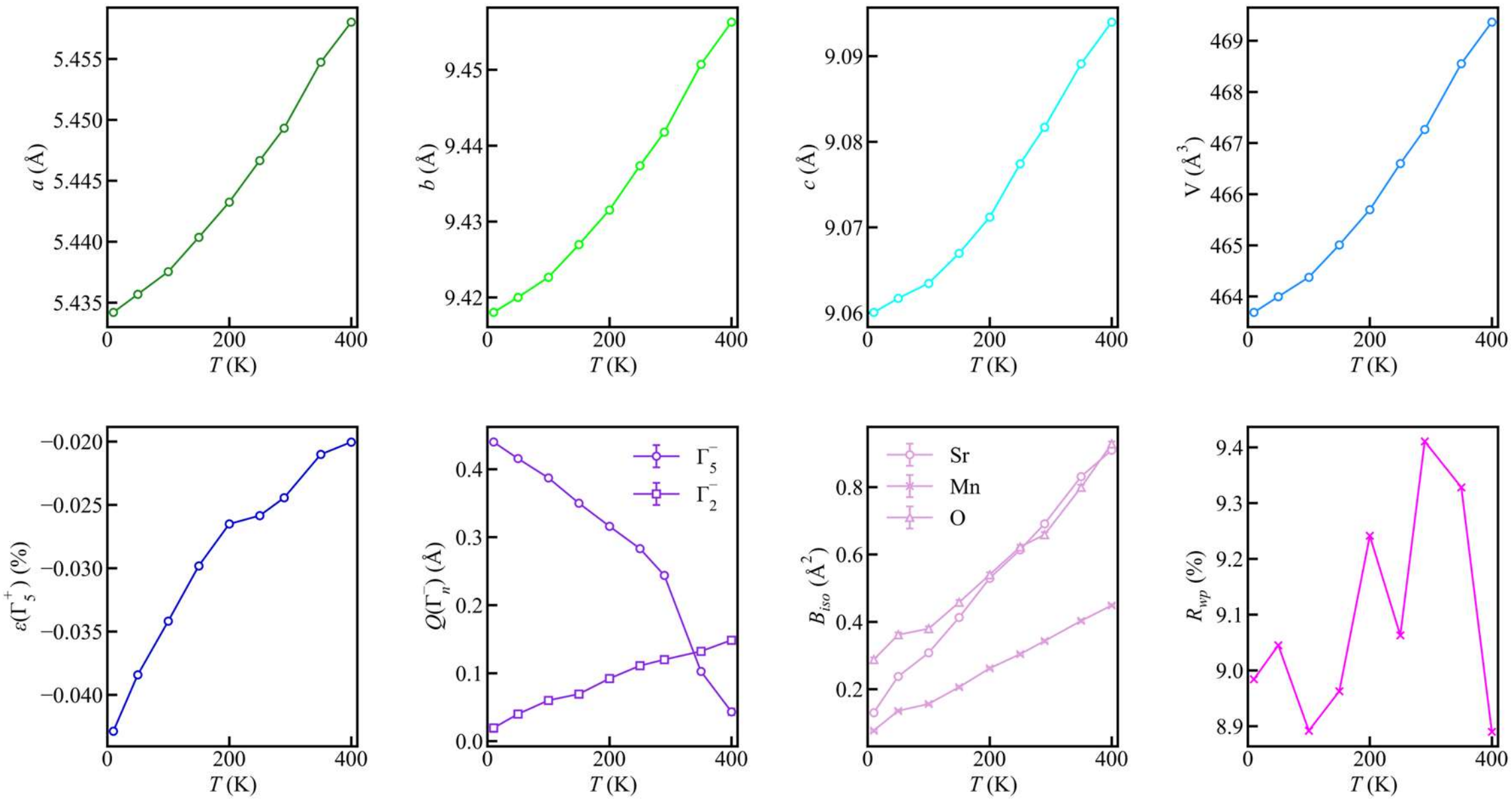

Figure S9. Selected fitting parameters from the final $Cmc2_1$ structural model refined against S-XRD data collected on ID22 (ESRF, France) in the range 10–400 K. Where not apparent, the error bars are smaller than the data points. For these refinements, only isotropic displacement parameters were refined to minimize the total number of refined parameters. The total $\Gamma_5^-$ and $\Gamma_2^-$ mode amplitudes ($Q$) were calculated based on the unconstrained mode components ($q_i$) as $Q = (\Sigma q_i^2)^{1/2}$. Note the apparent increase in $Q(\Gamma_2^-)$ at higher $T$ is an artefact of the refinements due to the use of isotropic displacement parameters.

Table S2. The $\Gamma_5^-$ mode components from the $Cmc2_1$ structural model refined against S-XRD data collected at 10 K. The relative weights of each component have been expressed as a percentage of the total mode amplitude, and the maximum displacement on each site ($d_{max}$) has been shown for reference. The O(1) and O(2) sites – corresponding to the oxygen atoms at the corner-sharing and face-sharing positions, respectively – form the dominant contribution to the total mode amplitude.

| Site | Symmetry | $Q(\Gamma_5^-)$ (Å) | Weight (%) | $d_{max}$ (Å) |
|---|---|---|---|---|
| Sr(1) | $E_u$ | 0.09653 | 5.5 | 0.06826 |
| Mn | E | 0.12837 | 9.8 | 0.06418 |
| O(1) | $A_u$ | 0.22994 | 31.3 | 0.11497 |
| | $B_{1u}$ | –0.24076 | 34.3 | 0.13900 |
| | $B_{2u}$ | 0.13036 | 10.1 | 0.07526 |
| O(2) | $B_1$ | –0.12300 | 9.0 | 0.07101 |
| Total | | 0.41080 | 100.0 | |

Table S3. Selected bond lengths obtained from Rietveld fits against the 10 K S-XRD data using the $Cmc2_1$ structural model.

| Sites | Distance (Å) | Sites | Distance (Å) |
|---|---|---|---|
| Mn(1)–O(1) | 1.852(2) x 2 | Mn(2)–O(1) | 1.882(2) x 1 |
| | 1.871(2) x 1 | | 1.899(2) x 2 |
| Mn(1)–O(2) | 1.863(2) x 2 | Mn(2)–O(2) | 1.904(3) x 1 |
| | 1.875(3) x 1 | | 1.946(2) x 2 |
| <Mn(1)–O> | 1.86(1) x 6 | <Mn(2)–O> | 1.91(2) x 6 |
| Mn(1)–Mn(2) | 2.4866(2) x 1 | | |

## Supplementary Note 3. DFT calculations.

Phonon calculations for the aristotype $P6_3/mmc$ structure reveal that the lowest frequency phonon mode at the zone center corresponds to the $\Gamma_5^-$ irrep (Fig. 3a). Condensation of the doubly degenerate $\Gamma_5^-$ instability along different order parameter directions results in a piezoelectric $C222_1$ phase and two polar phases with $Cmc2_1$ and $P2_1$ symmetries. The next unstable phonon mode transforms as the $\Gamma_3^-$ irrep and corresponds to another piezoelectric phase with $P\bar{6}2c$ symmetry. This is in perfect agreement with our previous RUM analysis, which showed that the $\Gamma_3^-$ mode represents the only other Γ-point RUM in the 4H polytype [3]. No intrinsic $\Gamma_2^-$ polar instability is observed in the phonon spectrum, and it solely appears as a secondary order parameter in the $Cmc2_1$ and $P2_1$ structures when the $\Gamma_5^-$ mode is condensed. Total energy calculations (Table S4) further confirm that the $Cmc2_1$ phase has the lowest energy, in agreement with experiment, and exhibits a dynamically stable phonon dispersion (Figure 3b).

Our DFT data demonstrate that the ground-state $Cmc2_1$ structure is characterized by a non-zero polarization, with a tilt amplitude of $Q(\Gamma_2^-) \approx 0.81$ Å and a polar mode amplitude of $Q(\Gamma_2^-) \approx 0.04$ Å. Under the point-charge approximation, the polar amplitude corresponds to a ground-state polarization of $P \approx 0.12$ μC cm$^{-2}$, which is comparable to the magnitude of spontaneous polarization found in other bulk improper ferroelectrics such as $TbMnO_3$ [4]. Polarization calculation using the Berry phase method results in a smaller value of ≈ 0.0245 μC cm$^{-2}$. Nevertheless, as first-principles calculations tend to mismatch the observed magnitude of the polarization in improper ferroelectrics, we used the $\Gamma_2^-$ components from the DFT-relaxed structure to estimate the experimental polarization based on our high-resolution S-XRD data. It can be extremely challenging to model polar distortions in improper ferroelectrics by Rietveld refinement alone as the weak nature of the polar distortions commonly effects only minor intensity differences to the existing Bragg reflections of the parent structure, hence polar mode amplitudes tend to be strongly correlated to other refined parameters. In the case of 4H-$SrMnO_3$, the freely refined polar mode amplitudes consistently increased with increasing temperature (Figure S9), reflecting their strong correlation to the (anisotropic) displacement parameters on the Sr(1) site. For this reason, we can only reliably model the polar distortion at 10 K where correlations between the refined polar mode amplitude and the displacement parameters will be minimal. By constraining our structural model in terms of the individual mode components extracted from the relaxed DFT structure, we obtained $Q(\Gamma_5^-) \approx 0.442(1)$ Å and $Q(\Gamma_2^-) \approx 0.018(8)$ Å. Both are approximately 50% of the DFT-relaxed amplitudes, showing relatively good agreement between the experimental and theoretical mode amplitudes.

To determine the magnetic ground state from DFT, we consider three different spin configurations: ferromagnetic (FM), A-type antiferromagnetic (A-AFM), and collinear ↑↑↓↓ AFM orderings, as shown in Figure S10. Our calculations confirm that the A-AFM ordering has the lowest energy, lying 54.737

meV/f.u. and 3.560 meV/f.u. below the FM and ↑↑↓↓ AFM orderings, respectively, in agreement with experimental observations and previous DFT studies [5].

As discussed in the main text, the wFM moment obtained for 4H-$SrMnO_3$ remains very small due to its insulating nature and the absence of any charge ordering on the Mn sites. To enhance the wFM instability and provide a microscopic route toward stabilizing the wFiM state, we performed charged DFT calculations by adding 0.1 electrons per Mn atom, resulting in a finite imbalance between the magnetic moments of the two inequivalent Mn sites, as shown in Table S6.

Table S4. Relative energies and mode details of DFT-relaxed structures of 4H-$SrMnO_3$. Relative energies ($\Delta E$) have been reported relative to the lowest-energy structure.

| Phase | $P6_3/mmc$ | $P6_3mc$ | $C222_1$ | $Cmc2_1$ | $P2_1$ |
|---|---|---|---|---|---|
| $\Delta E$ (meV/f.u.) | 14.802 | 13.140 | 0.092 | 0.000 | 0.157 |
| Primary irrep (Å) | N/A | $\Gamma_3^-$ ($a$): 0.35376 | $\Gamma_5^-$(a,0): 0.81656 | $\Gamma_5^-$(a,$\sqrt{3}$a/3): 0.81652 | $\Gamma_5^-$(a,b): 0.81746 |
| Secondary irreps (Å) | N/A | $\Gamma_1^+$(a): 0.00985 | $\Gamma_1^+$(a): 0.02213<br>$\Gamma_5^+$(a,$\sqrt{3}$a): 0.07289<br>$\Gamma_1^-$(a): 0.00998 | $\Gamma_1^+$(a): 0.02215<br>$\Gamma_5^+$(a,0): 0.06397<br>$\Gamma_2^-$(a): 0.04041 | $\Gamma_1^+$(a): 0.02231<br>$\Gamma_2^+$(a): 0.00004<br>$\Gamma_5^+$(a,b): 0.07180<br>$\Gamma_1^-$(a): 0.00888<br>$\Gamma_2^-$(a): 0.01321 |
| Strain modes | N/A | N/A | $\Gamma_5^+$(a,$\sqrt{3}$a): -0.00130 | $\Gamma_5^+$(a,0): -0.00163 | $\Gamma_5^+$(a,b):<br>a = -0.00027<br>b = 0.00133 |

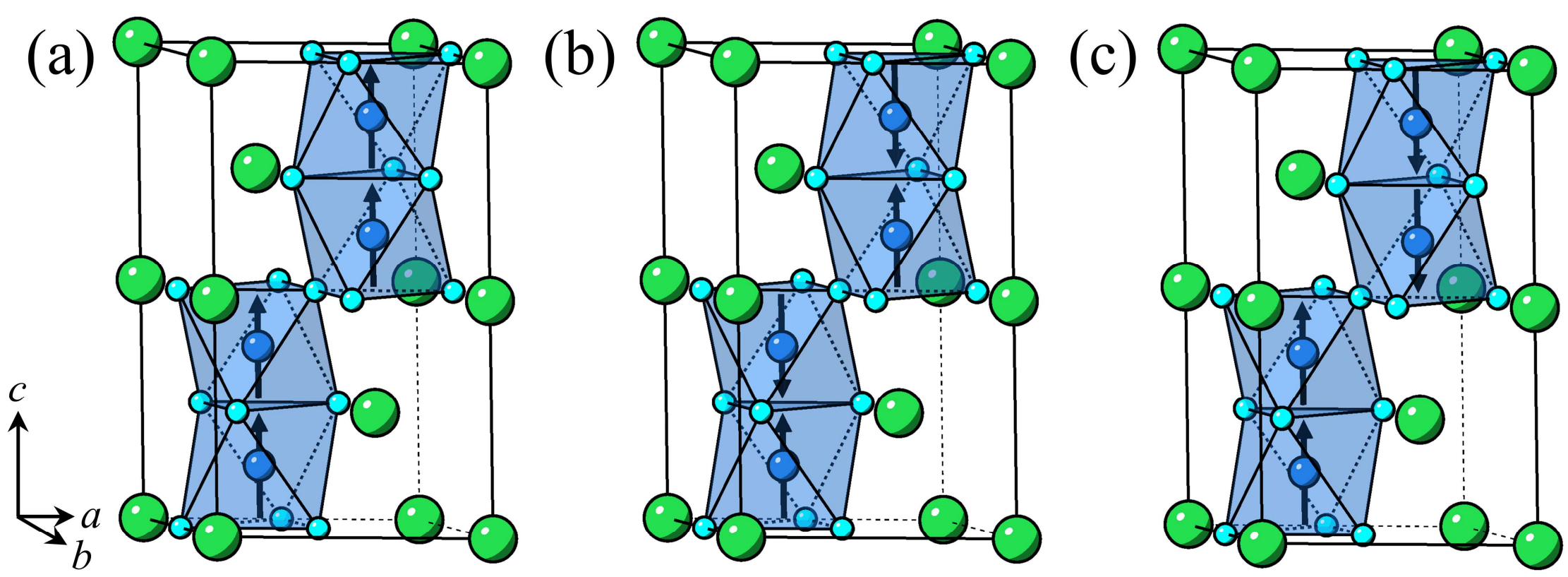

Figure S10. Collinear magnetic configurations considered in DFT calculations to determine the magnetic ground state: (a) Ferromagnetic (FM), (b) A-type antiferromagnetic (A-AFM), and (c) collinear ↑↑↓↓ AFM orderings. Green, blue, and cyan spheres correspond to Sr, Mn, and O atoms, respectively. Black arrows indicate the directions of Mn spin moments.

Table S5. Relative energies of the A-AFM ground state for different spin-axis orientations obtained from non-collinear DFT calculations including SOC with $U_{\text{eff}} = 4.5$ eV applied to the Mn 3d orbitals. $U_{\text{eff}}$ was varied within a reasonable range to verify the stability of the magnetic easy axis direction.

| Spin axis direction | $x$ | $y$ | $z$ |
|---|---|---|---|
| $\Delta E$ (meV/f.u.) | 0.000 | 0.007 | 0.073 |

Table S6. Magnetic mode details for the undoped and electron-doped phases in the A-AFM ground state with the magnetic easy-axis orientation. In the pristine phase, the two inequivalent Mn sites possess antiparallel magnetic moments with nearly identical magnitudes of 2.983 $\mu_B$ and 2.982 $\mu_B$, yielding a very small wFM moment. On the other hand, electron doping by 0.1 electrons per Mn induces a finite imbalance between the magnitudes of the antiparallel Mn moments, with values of 3.156 $\mu_B$ and 2.937 $\mu_B$, resulting in a finite amplitude of the wFM mode.

| Phase | Undoped (insulating) | Electron doped (weakly metallic) |
|---|---|---|
| Primary magnetic irrep ($\mu_B$) | $m\Gamma_6^-$ (a,0): 5.97043 | $m\Gamma_6^-$ (a,0): 6.09855 |
| Secondary magnetic irrep ($\mu_B$) | $m\Gamma_6^+$ (0,a): 0.00100 | $m\Gamma_6^+$ (0,a): 0.21920 |

## Supplementary Note 4. RUM analysis of the $\Gamma_5^-$ distortion.

Our present work is informed by our prior enumeration of the rigid-unit modes (RUMs) in various polytypes of the perovskite structure [3] using the ISOTILT webtool [6]. This tool can also be used to generate the list of atomic positions which are produced by each RUM, and the corresponding distorted structure can be decomposed in terms of symmetry-adapted displacement modes of the parent structure and compared against an observed structural distortion to determine whether it is consistent with the predicted displacement pattern of a given RUM. This allows RUM-based contributions to be distinguished from extraneous degrees of freedom (e.g., cation/anion displacements, Jahn-Teller distortions, strain) which transform under the basis of the same irrep. Normally, this information is trivial to discern in simple structure types such as 3C perovskites whose RUMs can be readily orthogonalized from extraneous distortions by virtue of a symmetry-adapted basis alone. However, for more complex structures, RUMs may transform as a linear combination of several distinct displacement modes acting on distinct Wyckoff sites which nonetheless collectively transform under the same irrep, so this irrep encompasses more degrees of freedom than are strictly necessary to reproduce a RUM-type displacement pattern. For these reasons, we sought to develop a simple workflow which could discriminate the RUM contribution to an observed structural distortion.

First, we provide some brief context to the RUM analysis performed by ISOTILT. In summary, ISOTILT constructs a series of symmetry-mode blocks which encompass all of the symmetry-allowed rotational modes of a given superstructure in terms of their irrep transformations with respect to the parent structure. These blocks contain information about how the pivot and shared atoms comprising a given framework transform under each irrep. A singular-value decomposition is then performed on this system of equations, allowing its null space to be analyzed to pinpoint the rotational modes which avoid internal distortions of the rigid units within the parent structure. This yields a series of right-singular vectors (RSVs) whose components delineate linear combinations of symmetry modes encompassing the displacement pattern of a potential RUM; the actual RUM character of each RSV can be assessed by calculating a relative root-mean-squared deviation (RrmsD) of the RSV components. A non-zero RrmsD characterizes an RSV with non-RUM character, while a zero (or negligible) RrmsD characterizes a pure (or quasi-) RUM. For reference, we provide example outputs for the $\Gamma_5^-$ symmetry-mode blocks in the 4H and 6H polytypes in Table S7, and we show representative structures of the RSVs with the lowest RrmsDs for the two polytypes in Figure S11.

Next, we outline our overall workflow to decompose a given structural distortion in terms of the RSVs identified by ISOTILT. ISOTILT can generate the list of atomic positions produced by the coefficients of a given RSV. These positions can then be used to construct a cif corresponding to a representative structure for each RSV. Using ISODISTORT, the RSV cifs can then be decomposed in terms of the displacive components of the pivot and passenger atoms (e.g., the atoms occupying the *B* and *X* sites, respectively, in an $ABX_3$ perovskite). The mode details for each RSV can then be exported, providing a basis from which to compare each RSV against an observed structural distortion. For this, we express the observed distortion $\mathbf{Q}_{\text{obs}}$ in terms of a linear combination of RSVs:

$$\mathbf{Q}_{\text{obs}} = \mathbf{Q}_0 + \sum_n c_n \mathbf{R}_n \quad (1)$$

where $\mathbf{R}_n$ are the RSVs, $c_n$ are coefficients representing the weight of each RUM, and $\mathbf{Q}_0$ represents the residual structural distortion transforming under the same irrep but not spanned by the RSV basis. The coefficients $c_n$ can be obtained through a least-squares fitting of Eq. (1) against a given $\mathbf{Q}_{\text{obs}}$ to decompose the contribution from each RSV. Although not all RSVs necessarily correspond to physically plausible distortions, including the full set of RSVs in the fitting routine provides an unbiased method to determine which RSVs best reproduce the observed distortion. This means we make no prior assumptions based on the RUM character of a given RSV to characterize the observed distortion.

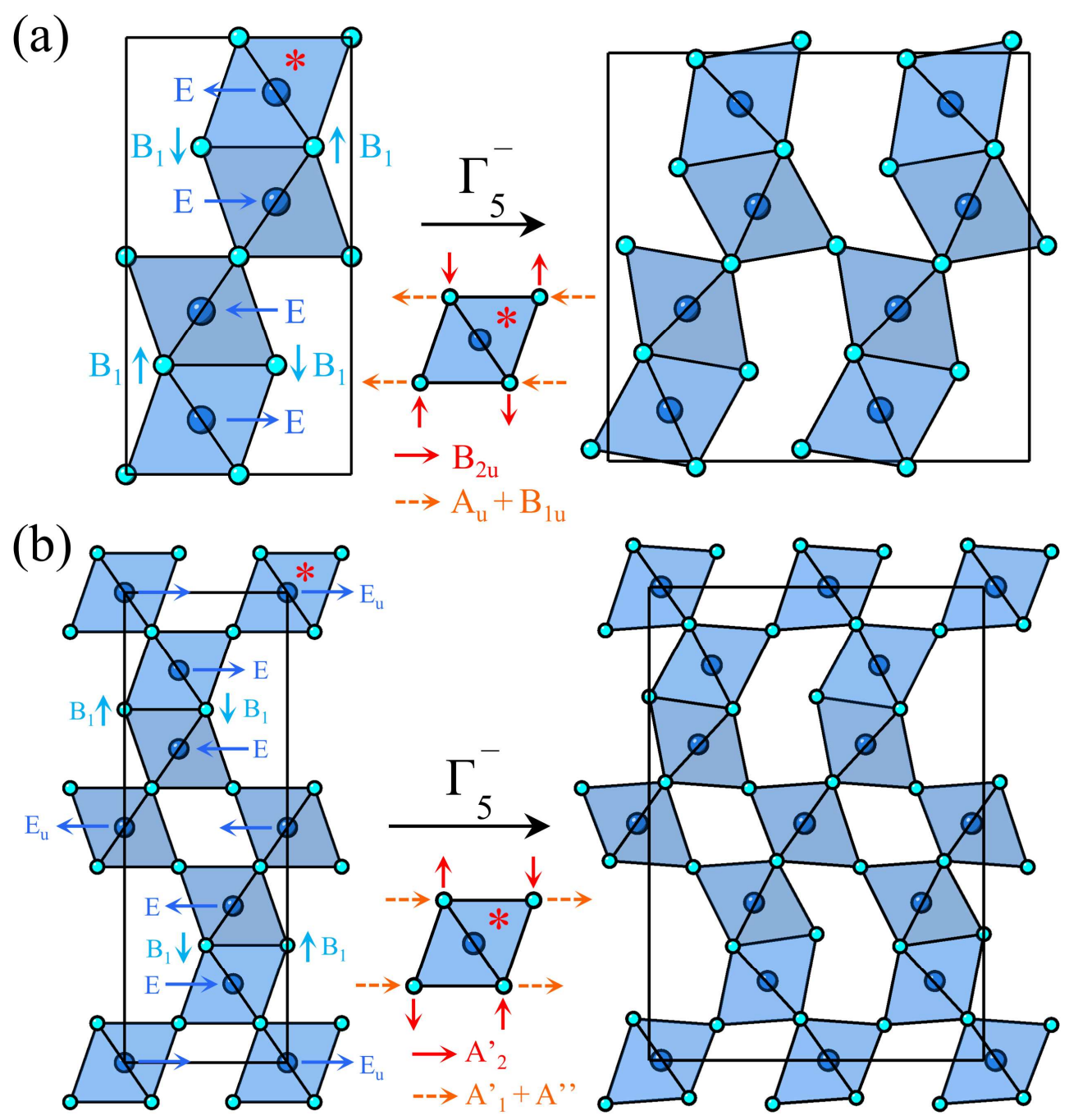

Figure S11. Displacement mode vectors for representative $\Gamma_5^-$ RSVs in the 4H (a) and 6H (b) polytypes. The irreps for the local Wyckoff positions have been labelled, and the corresponding point groups are provided in Tables S8 and S9. The vectors for the corner-sharing O(1) sites have been shown separately for visual clarity, with red asterisks denoting which octahedra the vectors have been depicted with respect to. Representative structures for each transformation been generated based on the RSVs output by ISOTILT; only the RSVs with the lowest relative root-mean-squared deviation (RrmsD) have been shown for simplicity.

To illustrate our approach, we first consider the experimental mode amplitudes we extracted from our S-XRD refinements for 4H-$SrMnO_3$ (Table S2). Table S8 summarizes the results of our RSV fitting against these mode amplitudes. Each RSV was renormalized based on the total mode amplitude of the observed distortion calculated without the displacement contribution from the Sr atoms – this is because the Sr atoms are not part of the polyhedral framework, hence they do not form part of any RUM-type motion. We find that the first RSV in Table S7 (**R**$_1$) forms the main contribution to the observed amplitudes ($c_1 \approx 1.04$), while the contribution from **R**$_2$ is significantly smaller ($c_2 \approx 0.09$). The residual structural distortion is almost entirely weighted towards the non-RUM Sr displacements (which are ~92% of the total residual amplitude), showing **R**$_1$ almost fully captures the observed distortion. Indeed, fitting **R**$_1$ alone gives a coefficient of $c_1 = 0.994414$ and reproduces ~98.9% of the observed distortion. This is the RSV which possesses RUM character, hence the RUM is by far the predominant driving structural instability in the parent structure of 4H-$SrMnO_3$. We consider this fitting method to be useful to discriminate the RUM character of structural distortions in other materials which transform as linear combinations of multiple displacive and rotational components.

Table S7. RUM analysis of the $\Gamma_5^-$ symmetry-mode blocks for the 4H and 6H polytypes. Each potential RUM consists of right-singular vectors (RSVs) whose components delineate the coefficients of linear combinations of displacive (dsp) and rotational (rot) symmetry modes. The relative root-mean-squared deviations (RrmsDs) of each RSV have been provided at the bottom of each column to distinguish their RUM character (RrmsD = 0).

| 4H polytype | | | | |
|---|---|---|---|---|
| Mode label | Site symmetry | Irrep | **R**$_1$ | **R**$_2$ |
| Mn1:f:dsp | $3m.$ | E | –0.76822 | 0.64019 |
| Mn1:f:rot | $3m.$ | E | –0.64019 | –0.76822 |
| | | RrmsDs | 0.00000 | 0.76878 |

| 6H polytype | | | | | |
|---|---|---|---|---|---|
| Mode label | Site symmetry | Irrep | **R**$_1$ | **R**$_2$ | **R**$_3$ |
| Ti1:a:dsp | $\bar{3}m.$ | $E_u$ | 0.73745 | –0.57737 | 0.35044 |
| Ti2:f:dsp | $3m.$ | E | 0.52141 | 0.81648 | 0.24796 |
| Ti2:f:rot | $3m.$ | E | 0.42929 | 0.00014 | –0.90317 |
| | | RrmsDs | 0.21484 | 1.00000 | 0.80734 |

For comparison, we show how this method may be applied to a non-RUM distortion. We use the 6H polytype of $BaTiO_3$, in which the $\Gamma_5^-$ mode stems from a SOJT distortion. The RSVs for this polytype are shown in Table S7, and the corresponding mode amplitudes and fitting results are shown in Table S9. In this case, the majority of the observed distortion is described by **R**$_2$, with the residual being largely weighted towards the Ba displacements (~80% of the total residual amplitude). In this case, **R**$_2$ only accounts for the displacements of the $Ti^{4+}$ cations, so its large contribution to the observed distortion is consistent with how SOJT distortions typically manifest in octahedral environments. However, Table S9 shows that the fitted distortion does not fully reproduce the observed distortion in this system, with the residual components for the O sites being of a similar order of magnitude to those of the observed distortion. This shows that the observed distortion in 6H-$BaTiO_3$ cannot be mapped purely in terms of the degrees of freedom spanned by the RSV basis. Again, this is consistent with how SOJT instabilities typically manifest in octahedral environments, where the Ti and O sites displace in a collinear fashion [7]. Overall, this demonstrates the sensitivity of our fitting approach to discriminate between RUM and non-RUM distortions.

Table S8. Mode amplitudes based on the $\Gamma_5^-$ RSVs in the 4H polytype predicted by ISOTILT ($\mathbf{R}_n$) compared to those obtained from the refinement against the ID22 data at 10 K ($\mathbf{Q}_{obs}$). The $\mathbf{R}_n$ components have been renormalized based on the total amplitude of $\mathbf{Q}_{obs}$ calculated without the Sr(1) component (i.e., $Q_{Tot-Sr}$ = 0.39930 Å), thus ensuring the ISOTILT mode amplitudes are comparable. The mode amplitudes reproduced by $\mathbf{R}_n$ ($\mathbf{Q}_{fit}$) as well as the residual amplitudes ($\Delta\mathbf{Q}$) are also provided.

| Site | Symmetry | Irrep | $\mathbf{R}_1$ (Å) | $\mathbf{R}_2$ (Å) | $\mathbf{Q}_{obs}$ (Å) | $\mathbf{Q}_{fit}$ (Å) | $\Delta\mathbf{Q}$(Å) |
|---|---|---|---|---|---|---|---|
| Sr(1) | $\bar{3}m.$ | $E_u$ | – | – | 0.09653 | 0.00000 | 0.09653 |
| Mn | $3m.$ | E | 0.12730 | 0.14905 | 0.12837 | 0.14675 | -0.01838 |
| O(1) | $.2/m.$ | $A_u$ | 0.22047 | -0.05677 | 0.22994 | 0.22474 | 0.00520 |
| | | $B_{1u}$ | -0.22047 | 0.05677 | -0.24076 | -0.22474 | -0.01602 |
| | | $B_{2u}$ | 0.15170 | -0.25571 | 0.13036 | 0.13439 | -0.00403 |
| O(2) | $mm2$ | $B_1$ | -0.15170 | 0.25571 | -0.12300 | -0.13439 | 0.01139 |
| | | $c_n$ | 1.04342 | 0.09343 | | | |

Table S9. Mode amplitudes based on the $\Gamma_5^-$ RSVs in the 6H polytype predicted by ISOTILT ($\mathbf{R}_n$) compared to those obtained experimentally in 6H-$BaTiO_3$ at 20 K ($\mathbf{Q}_{obs}$). The experimental mode amplitudes for 6H-$BaTiO_3$ have been taken from Ref. [2]. The $\mathbf{R}_n$ components have been renormalized based on the total amplitude of $\mathbf{Q}_{obs}$ calculated without the Ba components (i.e., $Q_{Tot-Ba}$ = 0.29979 Å). The mode amplitudes reproduced by $\mathbf{R}_n$ ($\mathbf{Q}_{fit}$) as well as the residual amplitudes ($\Delta\mathbf{Q}$) are also provided.

| Site | Symmetry | Irrep | $\mathbf{R}_1$ (Å) | $\mathbf{R}_2$ (Å) | $\mathbf{R}_3$ (Å) | $\mathbf{Q}_{obs}$ (Å) | $\mathbf{Q}_{fit}$ (Å) | $\Delta\mathbf{Q}$ (Å) |
|---|---|---|---|---|---|---|---|---|
| Ba(2) | $3m.$ | $E_u$ | – | – | – | 0.06846 | 0.00000 | 0.06846 |
| Ti(1) | $\bar{3}m.$ | $E_u$ | -0.10311 | -0.17309 | -0.06562 | -0.17260 | -0.17815 | 0.00555 |
| Ti(2) | $3m.$ | E | -0.07291 | 0.24477 | -0.04640 | 0.23516 | 0.23124 | 0.00392 |
| O(1) | $mm2$ | $B_1$ | -0.08522 | 0.00000 | 0.23259 | 0.06133 | 0.03817 | 0.02316 |
| O(2) | $.m.$ | $A'_1$ | 0.17752 | 0.00000 | -0.03365 | -0.02592 | -0.00685 | -0.01908 |
| | | $A'_2$ | 0.06028 | 0.00000 | -0.16447 | 0.00114 | -0.02699 | 0.02813 |
| | | A" | -0.17752 | 0.00000 | 0.03365 | -0.01863 | 0.00685 | -0.02548 |
| | | $c_n$ | -0.00801 | 0.97288 | 0.16116 | | | |

Supplementary Note 5. Variable-temperature S-XRD measurements.

We collected additional variable-temperature S-XRD data on the I11 beamline to study the evolution of the crystal structure across the structural and magnetic transitions. High-temperature data were collected using a Cyberstar hot air blower, while low-temperature data were collected using an Oxford Systems Cryostream. High- and low-temperature data were collected continuously using a Mythen position-sensitive detector while heating from 25 °C to 500 °C and 100 K to 300 K, respectively.

In a previous diffraction study [1], the exact transition temperature was noted to be difficult to ascertain due to the gradual suppression of the superstructure reflections near the phase transition. However, it was speculated that the structural transition likely coincided with a magnetization anomaly at $T_s = 380$ K so that this reflected the real transition temperature. However, our own S-XRD measurements show that $T_s$ exceeds 380 K (Fig. 2b), and fits of a simple Gaussian peak against the (261) superstructure reflection yield an estimate of $T_s \approx 458.4(8)$ K (Figure S12). This temperature is likely slightly overestimated as the peak begins to merge with the background near $T \approx 450$ K. Nevertheless, the FE distortion in this system should persist well above room temperature.

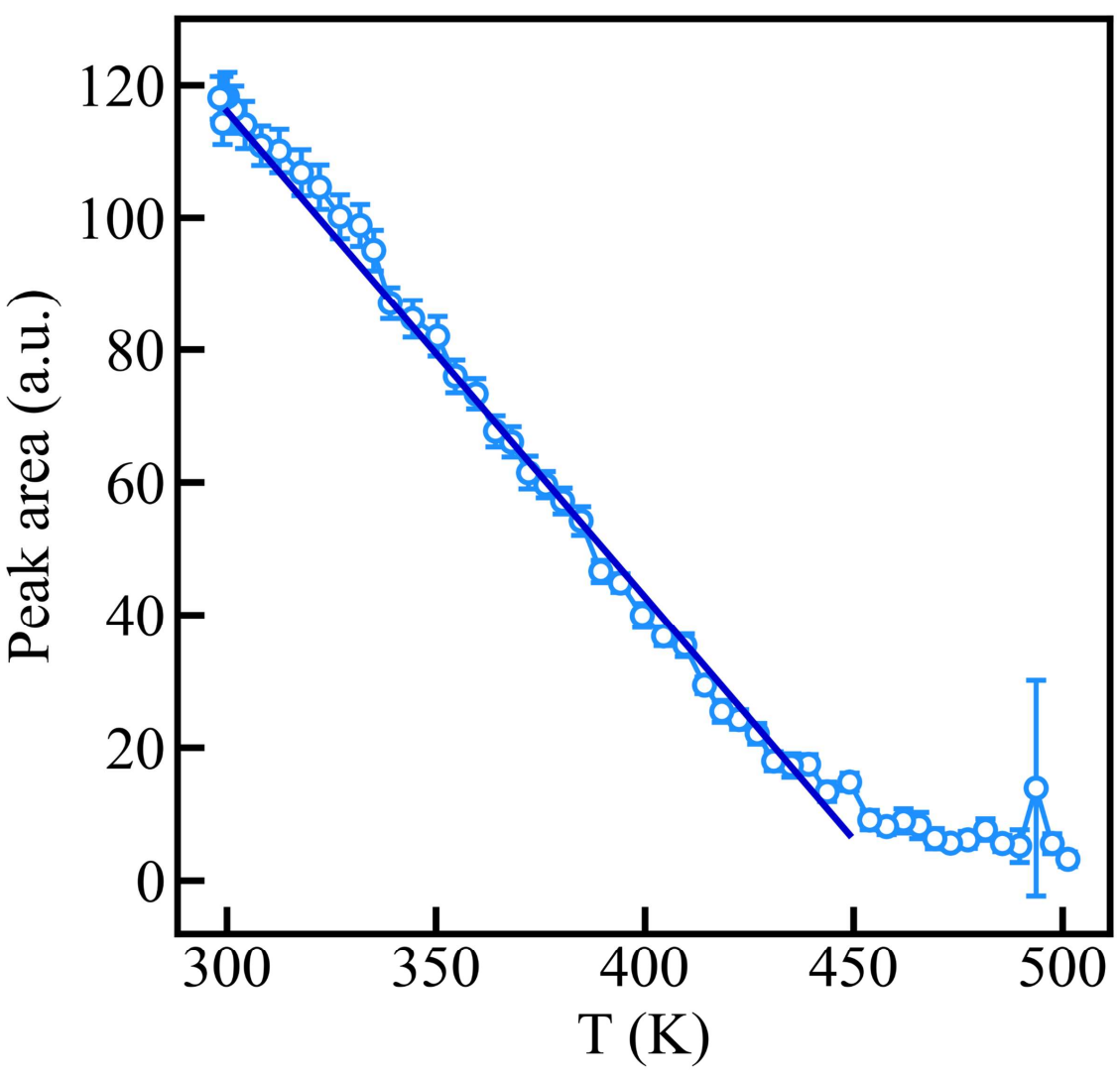

Figure S12. Areas of a Gaussian peak fitted against the (261) superstructure reflection in the $Cmc2_1$ phase. A linear regression was used to estimate the transition temperature as 459(1) K.

The fitted lattice parameters show no clear anomalies in the range 100–800 K, exhibiting positive thermal expansion throughout (Figure S13, Figure S14). These fits were based on a $Cmc2_1$ structure first modelled against the 300 K data, before performing sequential refinements against the lower- and higher-temperature datasets. Our recent diffraction study of 6H-$BaTiO_3$ showed its $Cmc2_1 \rightarrow C222_1$ phase transition is marked by a discontinuous change in lattice parameters [2], so the lack of any similar discontinuities in our current diffraction data strongly suggests no analogous structural transition occurs in 4H-$SrMnO_3$. The long-range AFM transition at $T_N$ also does not appear to be accompanied by any change in structural symmetry, though the volume derivative with respect to temperature reveals the magnetic phase transition is distinguished by a subtle magnetostrictive effect. Additional anomalies in the refined magnitudes of the ferroelastic strain ($\varepsilon(\Gamma_5^+)$) and the $\Gamma_5^-$ tilts ($Q(\Gamma_5^-)$) corroborate this. As we found for our ID22 data (Supplementary Note 2), the apparent increase in the polar mode amplitudes ($Q(\Gamma_2^-)$) at higher temperatures reflects strong correlations between the refined polar mode components and the anisotropic displacement parameters.

The magnetostriction at $T_N$ does not appear to significantly affect the local $Mn^{4+}$ coordination environments (Figure S15). The intradimer Mn–Mn distances exhibit an overall linear decrease in the range 100–300 K so that the appearance of long-range AFM order does not appear to trigger any notable change in the direct exchange interactions within the face-sharing bioctahedra. The concomitant anomalies in $Q(\Gamma_5^-)$ and $\varepsilon(\Gamma_5^+)$ at $T_N$ suggest the magnetostriction more likely reflects the onset of coupling between the tilts and the AFM/wFM orders, which is consistent with our symmetry predictions. Interestingly, the Mn–O2 distances appear to exhibit anomalous changes in slope at ~240 K. This does not appear to be related to the AFM ordering at $T_N$, and we detect no other anomalies at this temperature, so further work would be necessary to explore this in more detail.

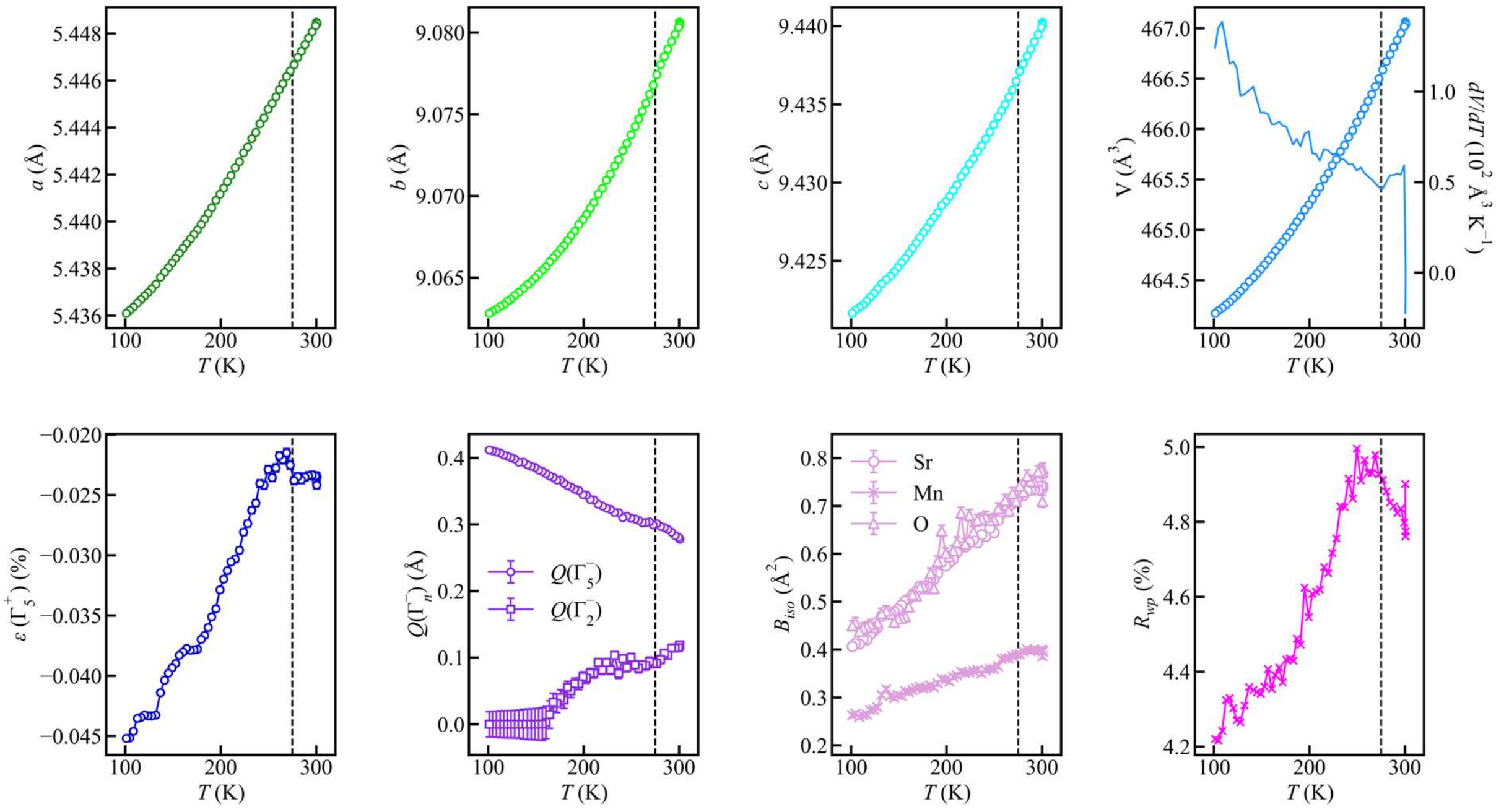

Figure S13. Selected fitting parameters from the $Cmc2_1$ structural model sequentially refined against S-XRD data collected on I11 (Diamond Light Source, U.K.) in the range 100–300 K. The dashed vertical lines denote the AFM transition ($T_N \approx 280$ K). The derivative of $V$ with respect to $T$ has also been shown to highlight a subtle magnetostrictive effect at $T_N$. The average temperature interval between data points was ~4 K. Where not apparent, the error bars are smaller than the data points.

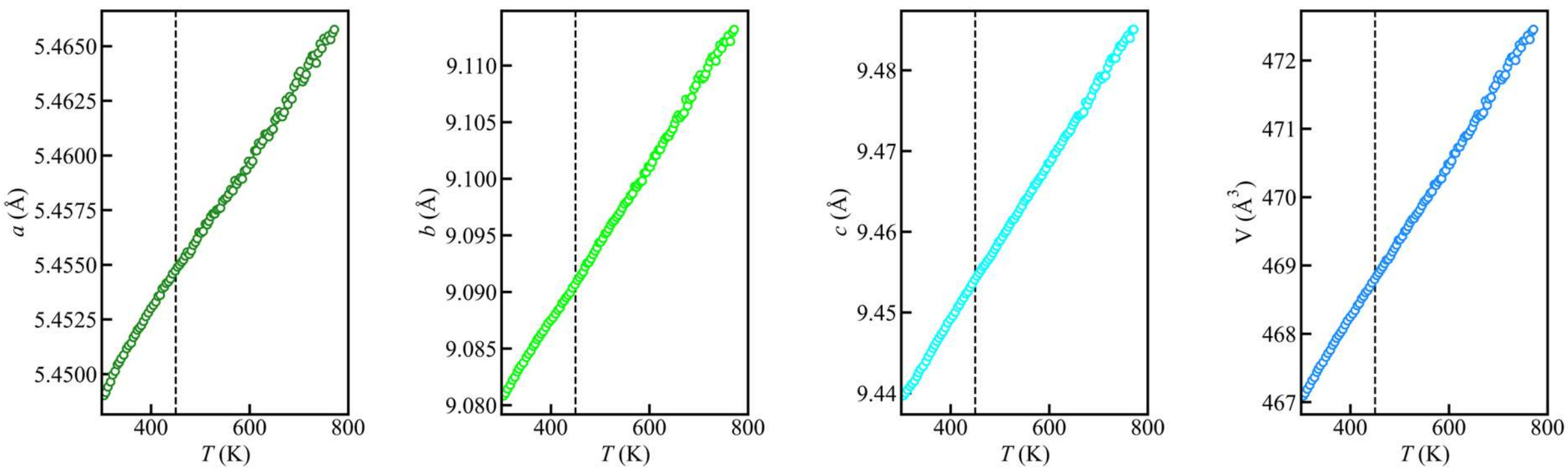

Figure S14. Selected fitting parameters from the $Cmc2_1$ structural model sequentially refined against S-XRD data collected on I11 (Diamond Light Source, U.K.) in the temperature range 300–800 K. The dashed vertical lines denote $T_S \approx 450$ K. As the structure transitions to $P6_3/mmc$ symmetry above $T_S$, only the variation in lattice parameters has been shown to highlight the lack of any additional phase transitions above room temperature. The average temperature interval between data points was ~4 K. Where not apparent, the error bars are smaller than the data points.

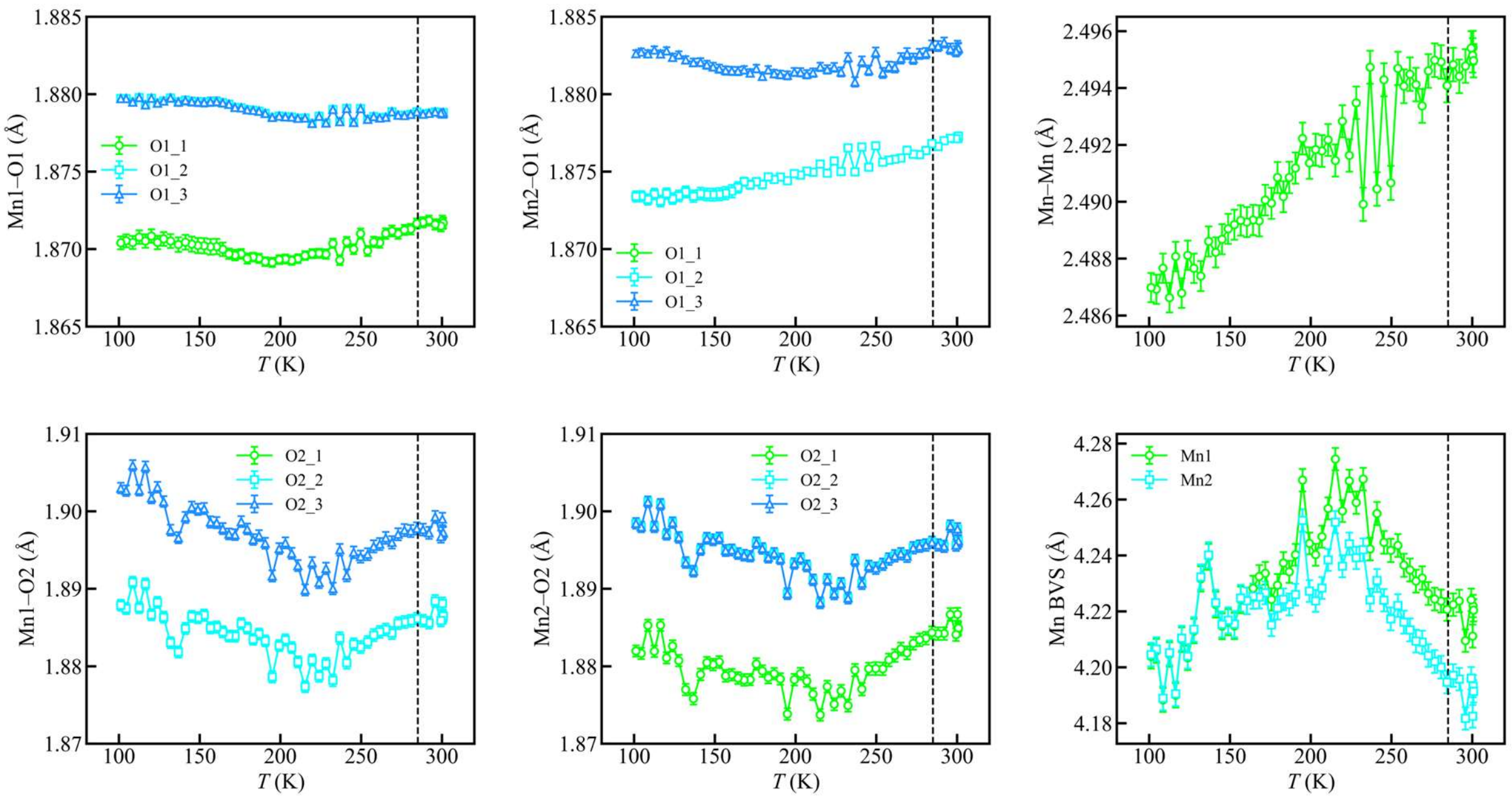

Figure S15. Selected interatomic distances and bond valence sums (BVS) based on sequential refinements of a $Cmc2_1$ model against low-temperature S-XRD data collected at I11 (Diamond Light Source, U.K.). The dashed vertical lines denote the AFM transition ($T_N \approx 280$ K). The O(1) and O(2) sites correspond to the oxygen atoms at the corner- and face-sharing positions, respectively.

Supplementary Note 6. Neutron diffraction measurements and magnetic structure analysis.

To investigate the magnetic structure of 4H-$SrMnO_3$, we performed NPD experiments using the GEM diffractometer at ISIS. Inspection of the room-temperature NPD patterns (Figure S16) revealed no traces of magnetic Bragg peaks, and the $Cmc2_1$ structural model obtained from the S-XRD refinements provided a good fit against these patterns (overall $R_{wp}$ = 5.24%). The Sr, Mn, and O sites could all be refined to within 1% of full occupancy, substantiating that the nominal stoichiometry is accurate (which was also corroborated by our S-XRD refinements). The final refined structural parameters and reliability factors obtained from the multi-bank refinements are provided in Table S10.

Table S10. Crystal and structural parameters extracted from the multi-bank GEM refinement based on data collected at 290 K.

| 4H-$SrMnO_3$, 290 K, SG: $Cmc2_1$, $Z$ = 8 | | | | |
|---|---|---|---|---|
| | $a$ (Å) | 5.4503(7) | | |
| | $b$ (Å) | 9.4411(1) | | |
| | $c$ (Å) | 9.0810(1) | | |
| | $V$ (Å$^3$) | 467.279(12) | | |
| | $R_p$ (%) | 5.5711 | | |
| | $R_{wp}$ (%) | 5.2269 | | |
| | GOF | 1.56 | | |
| Site | x | y | z | $B_{iso}$ (Å$^2$) |
| Sr1 | 0 | –0.0048(13) | 0 | 0.37(3) |
| Sr2 | 0 | 0.3309(23) | 0.2472(14) | 0.37(3) |
| Mn1 | 0 | 0.3385(19) | 0.6141(12) | 0.06(3) |
| Mn2 | 0 | 0.6716(19) | 0.3887(12) | 0.06(3) |
| O1_1 | 0 | 0.4941(17) | 0.0035(24) | 0.41(2) |
| O1_2 | 0.2495(23) | 0.2621(9) | 0.4941(16) | 0.41(2) |
| O2_1 | 0 | 0.8207(24) | 0.2474(16) | 0.41(2) |
| O2_2 | 0.7661(21) | 0.4068(11) | 0.7503(11) | 0.41(2) |

At 10 K, additional magnetic scattering contributions appear at low Q (see Fig. 3a in the main text). These scattering features can be modelled using the previously reported spin configuration consisting of a layered AFM arrangement [1]. Using ISODISTORT [8,9], we found that this magnetic structure transforms in terms of the parent $P6_3/mmc$ structure as the $m\Gamma_6^-$ irrep. As we are insensitive to the precise OPD in our NPD measurements, we used the (a,0) $m\Gamma_6^-$ OPD obtained from our DFT calculations (Table S5) as our model, corresponding to the Shubnikov group $Cmc'2_1'$. As we are insensitive to the $m\Gamma_6^-$ OPD, we only use this model to refine the overall Mn moment, rather than verify the spatial orientation of the spins. Rietveld fits using this model (Figure S17) yielded a refined Mn moment of 2.34(3) $\mu_B$ per Mn at 10 K – this agrees well with the moment of 2.27(1) $\mu_B$ obtained in the previous NPD study [1], which is reduced by ~25% from the ordered moment of 3 $\mu_B$ expected for high-spin $Mn^{4+}$ due to covalency effects [5]. Full crystallographic details are provided in Table S11.

Table S11. Crystal, structural, and magnetic parameters extracted from the multi-bank GEM refinement based on data collected at 10 K.

| 4H-$SrMnO_3$, 10 K, SG: *Cmc*'$2_1$', $Z = 8$ | | | | |
|---|---|---|---|---|
| | $a$ (Å) | 5.4352(3) | | |
| | $b$ (Å) | 9.4237(6) | | |
| | $c$ (Å) | 9.0621(1) | | |
| | $V$ (Å$^3$) | 464.157(13) | | |
| | $R_p$ (%) | 2.7172 | | |
| | $R_{wp}$ (%) | 3.4564 | | |
| | GOF | 1.76 | | |
| Site | x | y | z | $B_{iso}$ (Å$^2$) |
| Sr1 | 0 | –0.0080(5) | 0 | 0.00(2) |
| Sr2 | 0 | 0.3323(10) | 0.2566(14) | 0.00(2) |
| Mn1 | 0 | 0.3368(13) | 0.6124(17) | –0.01(3) |
| Mn2 | 0 | 0.6766(13) | 0.3868(17) | –0.01(3) |
| O1_1 | 0 | 0.4855(10) | 0.0139(18) | 0.20(2) |
| O1_2 | 0.2495(13) | 0.2652(5) | 0.4993(17) | 0.20(2) |
| O2_1 | 0 | 0.8194(11) | 0.2416(16) | 0.20(2) |
| O2_2 | 0.7650(11) | 0.4070(7) | 0.7551(15) | 0.20(2) |
| Site | $M_x$ | $M_y$ | $M_z$ | $M_{total}$ |
| Mn1 | 2.34(3) | 0 | 0 | 2.34(3) |
| Mn2 | –2.34(3) | 0 | 0 | 2.34(3) |

Unlike the $\Gamma_5^-$ OPD in our S-XRD experiments, we cannot use the $\Gamma_5^+$ ferroelastic strain to infer the m$\Gamma_6^-$ OPD as there is no unique mapping between the two bases spanned by these irreps (Figure S18). We also tested all possible magnetic space groups generated by the direct product of the m$\Gamma_6^-$ and $\Gamma_5^-$ irreps against our NPD data (Table S12), but we obtained nearly identical residuals in each case (overall $R_{wp} \approx 3.46\%$). For these reasons, we are insensitive to the precise magnetic space group assignment based on our existing NPD data alone. Nevertheless, we can still deduce the character of the wFM component based on symmetry arguments. Table S12 shows all possible magnetic irreps which are generated through the direct product of the $\Gamma_5^-$ and m$\Gamma_6^-$ irreps. No matter which $\Gamma_5^-$ and m$\Gamma_6^-$ OPDs are experimentally adopted, a secondary magnetic component transforming as m$\Gamma_6^+$ is always present, and this is permitted by the site symmetries of the 4H polytype to yield a non-zero net moment across the $Mn^{4+}$ sites. None of the other secondary magnetic representations obtained by the direct products of $\Gamma_5^-$ and m$\Gamma_6^-$ correspond to FM orders, and the m$\Gamma_6^-$ irrep alone is insufficient to yield the m$\Gamma_6^+$ component on an improper basis. These symmetry arguments highlight the essential role of the $\Gamma_5^-$ tilts to induce the wFM component in 4H-$SrMnO_3$. Indeed, this is precisely why no weak FM component emerges at $T_N$ in the untilted Ba analogue 4H-$BaMnO_3$ [10] despite this material sharing the same m$\Gamma_6^-$ AFM spin configuration in the ground state [11].

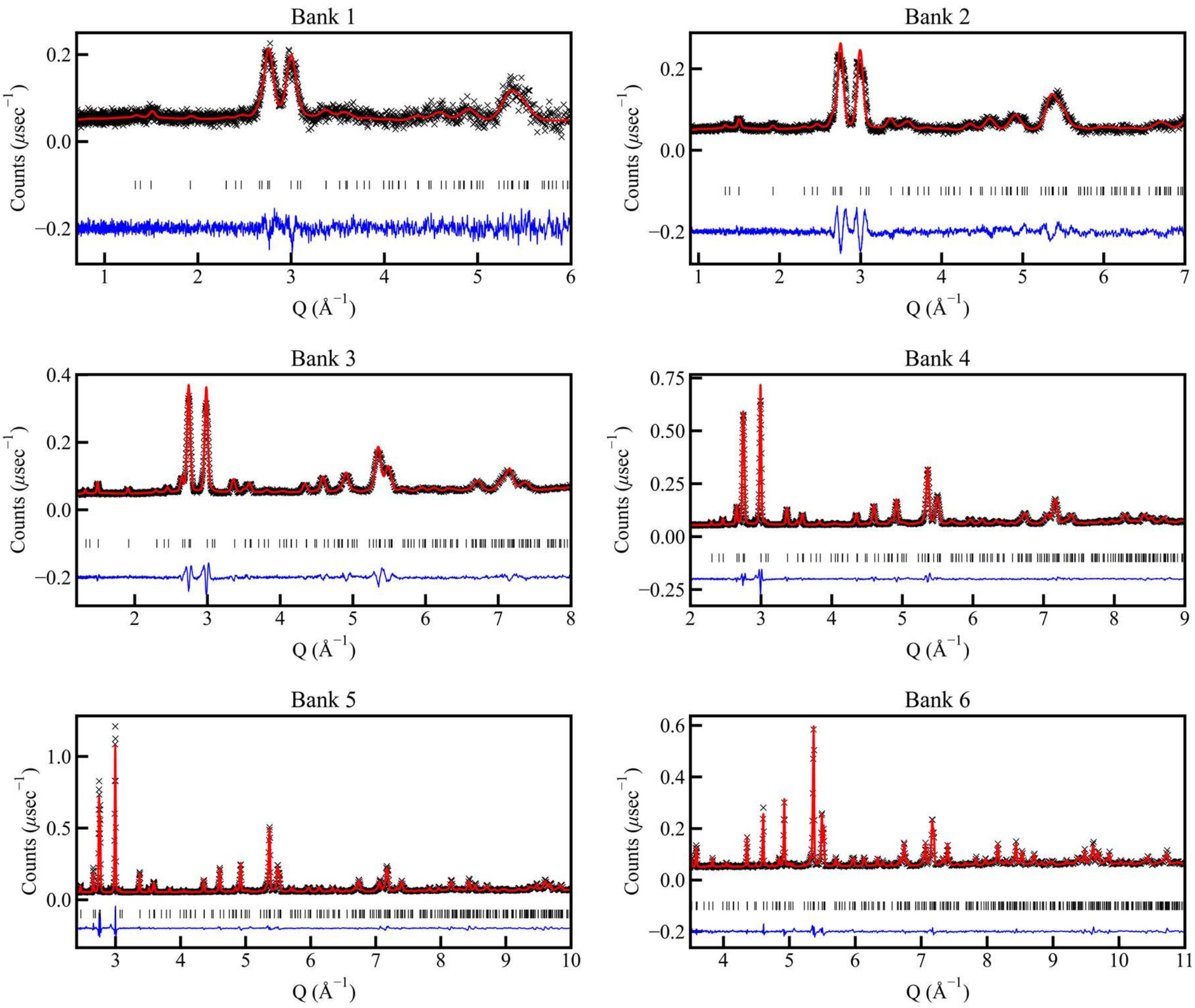

Figure S16. Multi-bank Rietveld fit of the $Cmc2_1$ structural model against room-temperature NPD data collected on GEM. Black ticks represent the positions of the nuclear Bragg reflections. Overall $R_{wp}$ = 5.24%.

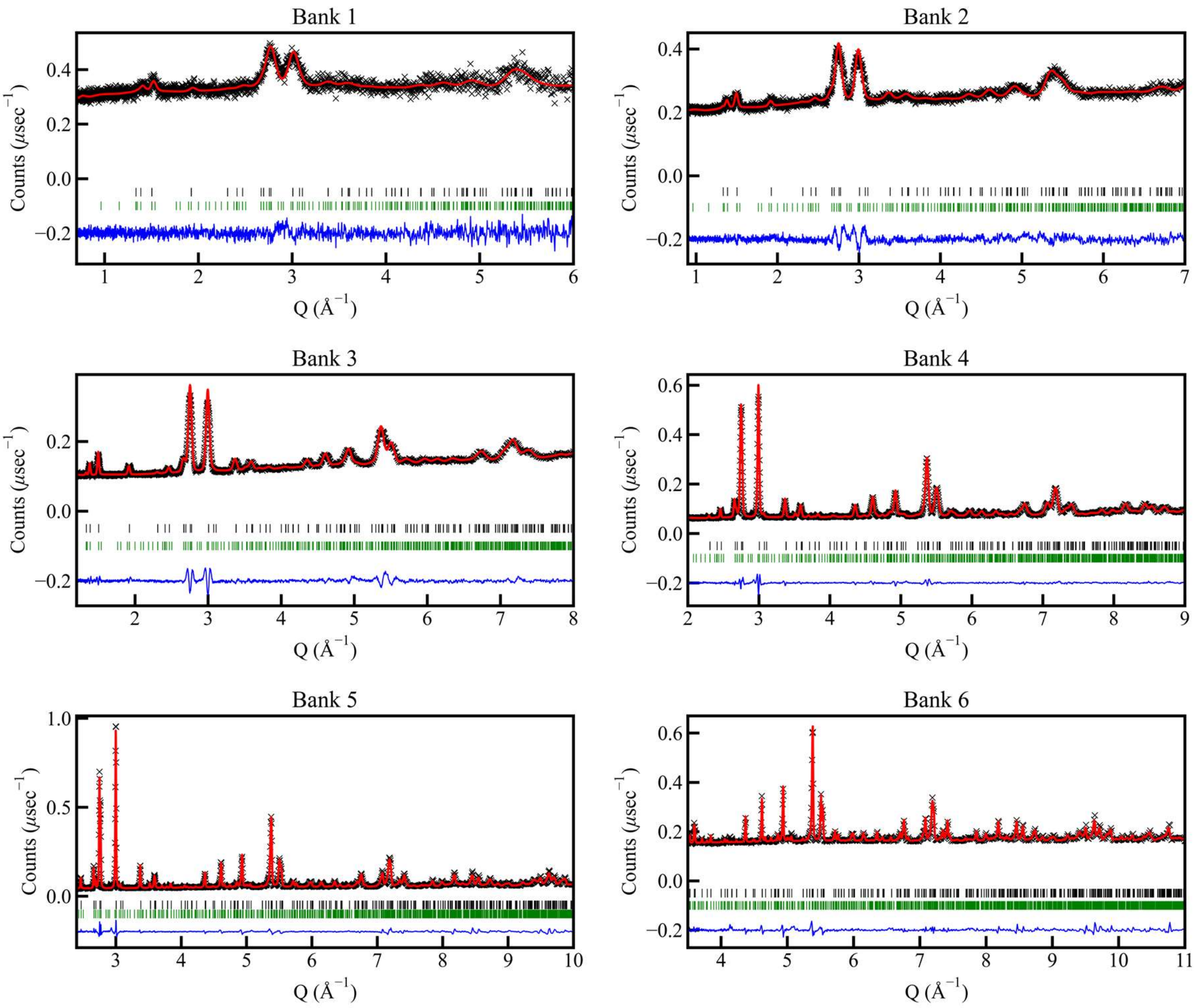

Figure S17. Multi-bank Rietveld fit of the *Cmc*'$2_1$' magnetic structural model against 10 K NPD data collected on GEM. Black and green ticks represent the positions of the nuclear and magnetic Bragg reflections, respectively. Overall $R_{wp}$ = 3.73%.

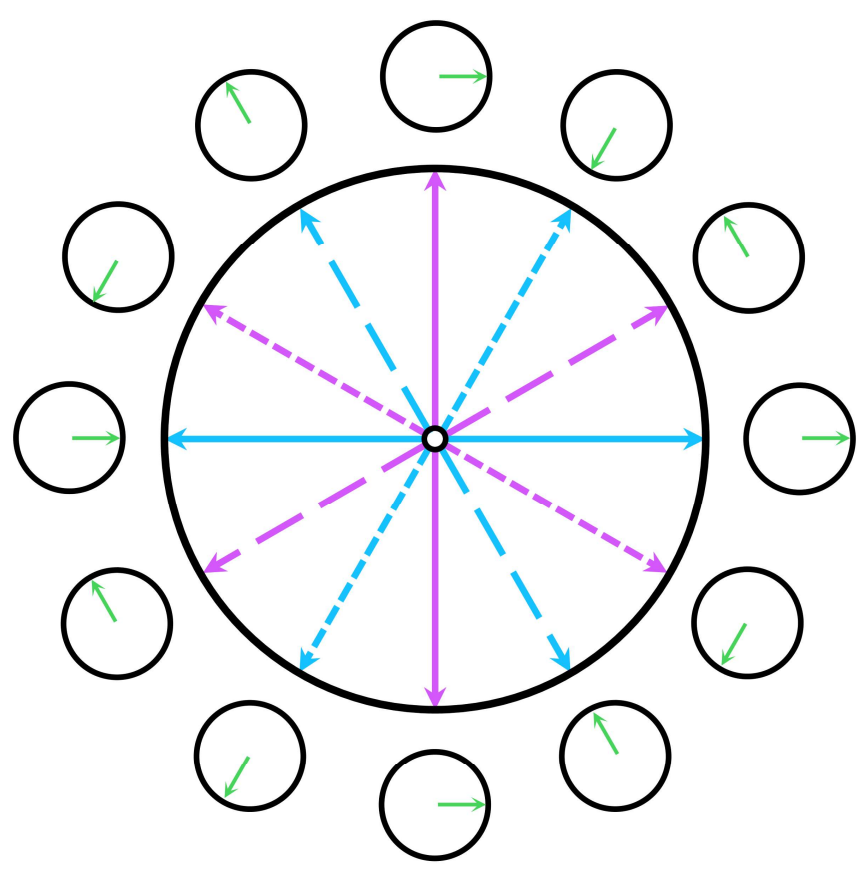

Figure S18. The propagation between the $m\Gamma_6^-$ and $\Gamma_5^+$ order parameter spaces, taking $m\Gamma_6^-$ as the primary order parameter. The central circle corresponds to the $m\Gamma_6^-$ space, with blue and magenta arrows denoting the two sets of high-symmetry OPDs corresponding to *Cmc'm* and *Cm'cm* symmetries, respectively. The outer circles show how the choice of $m\Gamma_6^-$ OPD propagates to the $\Gamma_5^+$ OPD. Arrows of the same line style in the central circle map to the same strain OPD, highlighting that equivalent strain OPDs may map to distinct $m\Gamma_6^-$ OPDs.

Table S12. All combinations of secondary magnetic irreps based on direct products of the primary $\Gamma_5^-$ and $m\Gamma_6^-$ irreps. Note that only the irreps which are permitted by the site symmetries of the 4H polytype to yield a non-zero moment have been shown. The epikernels of the primary irreps as well as the final Shubnikov groups obtained by the direct products of $\Gamma_5^-$ and $m\Gamma_6^-$ have been provided next to each OPD for reference. The only representation corresponding to ferromagnetic order is $m\Gamma_6^+$ (highlighted in bold), and this irrep is obtained through all possible direct products of $\Gamma_5^-$ and $m\Gamma_6^-$.

| | $\Gamma_5^-$ OPD | | |
|---|---|---|---|
| $m\Gamma_6^-$ OPD | (a,0) – $C222_1$ | (a, $\sqrt{3}$a/3) – $Cmc2_1$ | (a, b) – $P2_1$ |
| (a,0) – *Cmc'm* | $m\Gamma_4^+$, $\mathbf{m\Gamma_6^+}$ ($C22'2_1'$) | $\mathbf{m\Gamma_6^+}$ ($Cmc'2_1'$) | – |
| (a, $\sqrt{3}$a/3) – *Cm'cm* | $m\Gamma_3^-$, $\mathbf{m\Gamma_6^+}$ ($C22'2_1'$) | $m\Gamma_3^-$, $m\Gamma_4^+$, $\mathbf{m\Gamma_6^+}$ ($Cm'c2_1'$) | – |
| (a, b) – $P2_1'/m$ | – | – | $m\Gamma_3^-$, $m\Gamma_4^+$, $\mathbf{m\Gamma_6^+}$ ($P2_1'$) |

## Supplementary Note 7. Magnetic property measurements

Figure S19 shows full magnetic susceptibility data for both 4H-$BaMnO_3$ and 4H-$SrMnO_3$. The sample of 4H-$BaMnO_3$ was prepared following the same protocol for 4H-$SrMnO_3$, with the final heating step performed at 1325 °C instead of 1250 °C to avoid the formation of unwanted polytypes [11], and similar to our 4H-$SrMnO_3$ samples, only a trace amount of a $Mn_3O_4$ impurity phase was formed (~0.4 wt.% by Rietveld refinement). The susceptibility data for 4H-$BaMnO_3$ is broadly consistent with previous reports, with the zero-field-cooled (ZFC) and field-cooled (FC) data largely overlapping until ~45 K. This feature was previously proposed to represent a second magnetic transition involving a canting of the $Mn^{4+}$ spins [11], but our present symmetry insights on 4H-$SrMnO_3$ (namely, that the wFM moment is only generated through coupling to the $\Gamma_5^-$ tilts) refute this possibility. We note that the spinel-type $Mn_3O_4$ impurity phase detected in our diffraction experiments also displays a large ZFC-FC divergence below 42 K [12], so we attribute the low-temperature divergence in both samples to this impurity. We do not detect any other magnetic impurities in either our S-XRD or NPD measurements.

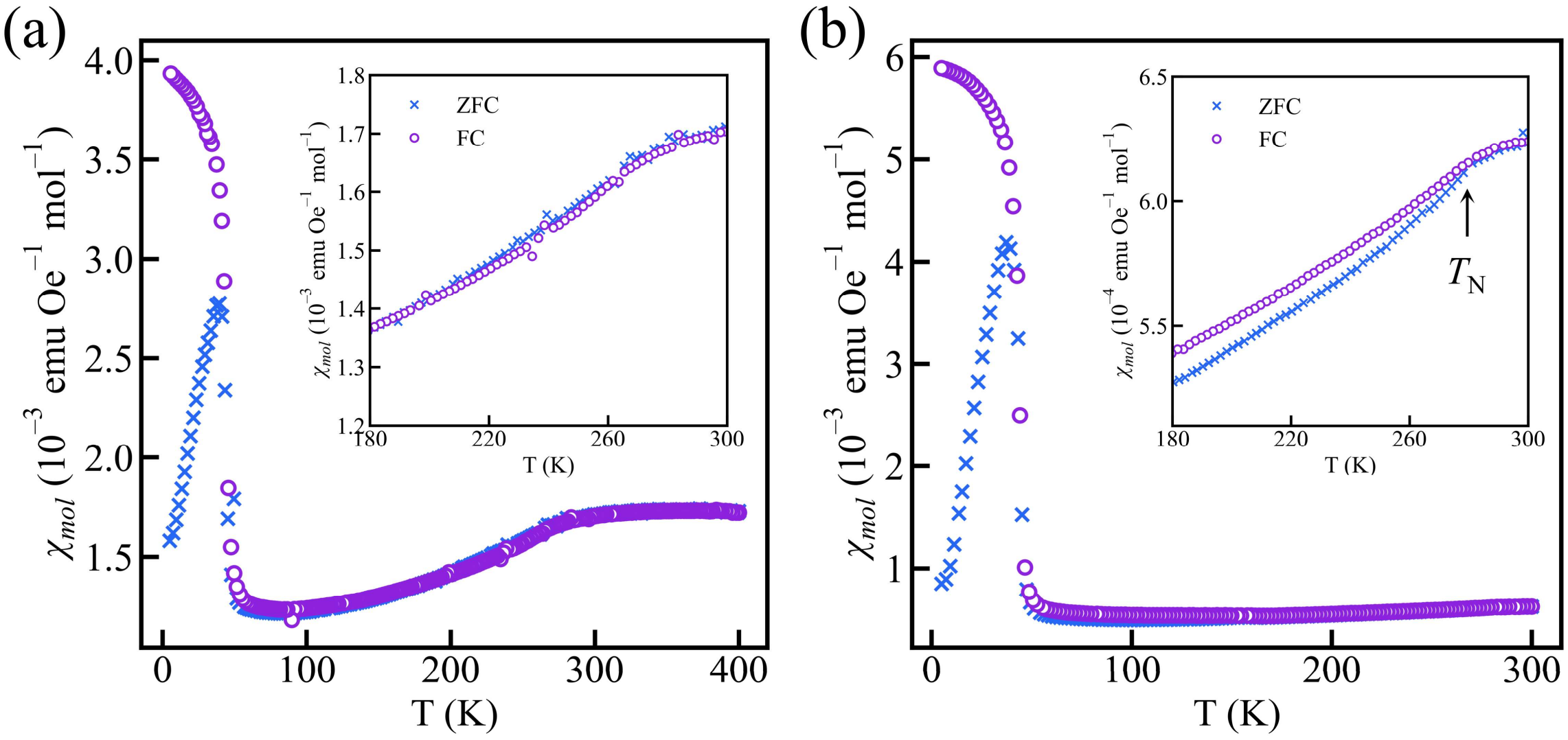

Figure S19. Comparison of the molar magnetic susceptibility ($\chi_{mol}$) for (a) 4H-$BaMnO_3$ and (b) 4H-$SrMnO_3$. The insets expand the region around the Néel temperature ($T_N$), highlighting the lack of divergence between the ZFC and FC traces for 4H-$BaMnO_3$.

The raw hysteresis loops for 4H-$SrMnO_3$ collected between 200 K and 300 K are shown in Figure S20a. The overall response is nearly linear, though consistent with previous reports [13] the low-field response (–2 kOe < $H$ <2 kOe) is non-linear, suggesting a possible spin-flop transition occurs. This likely also relates to an unusual anomaly detected in our DC resistivity data which was suppressed upon the application of a magnetic field (Supplementary Note 8). Careful inspection of the remanent magnetization ($M_r$) shows a systematic enhancement in the wFM component below $T_N$, reflecting coupling between the long-range AFM order and the tilts. A residual $M_r$ persists above $T_N$, but given that an extended regime of short-range AFM order was reported to precede $T_N$ [14,15], we conclude that this still reflects the tilt coupling.

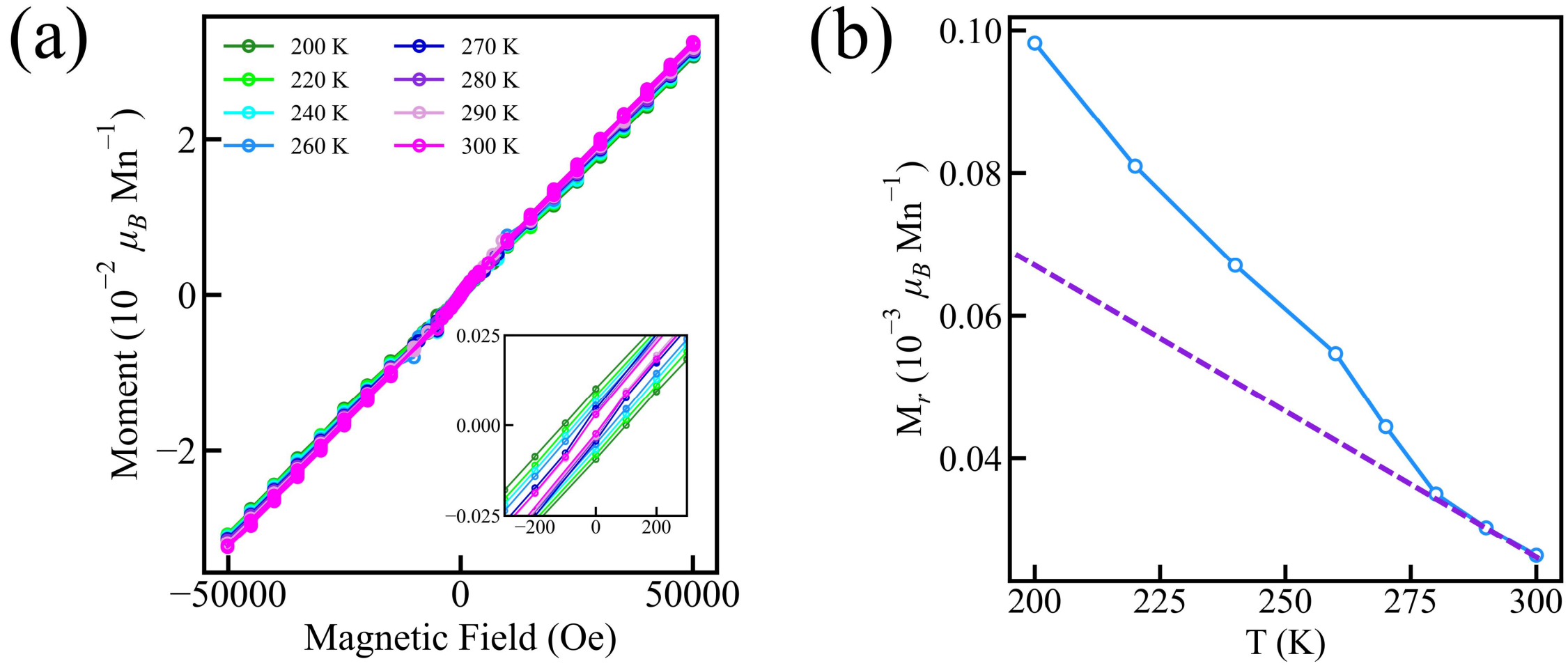

Figure S20. (a) Magnetization versus applied field loops collected between 200 K and 300 K for 4H-$SrMnO_3$. The inset expands the low-field regime to highlight the opening of a wFM component below $T_N$. (b) The change in remanent magnetization ($M_r$), highlighting the enhancement below $T_N \approx 280$ K.

Our charged DFT calculations suggested that electron-doped samples of $SrMnO_3$ would allow for the magnitude of the wFM moment to be enhanced. To verify this, we prepared a reduced sample of $SrMnO_{3-x}$ by heating oxygen-stoichiometric $SrMnO_3$ to 1250 °C for 24 h before quenching in air. By comparing the refined lattice parameters (Table S13) to reference literature values [16], the sample stoichiometry can be estimated as $SrMnO_{\sim 2.97}$, corresponding to ~0.06 $e^-$ per Mn. Subsequent DC susceptibility measurements (Figure S21) showed that the magnitude of the wFM moment was enhanced relative to the oxygen-stoichiometric sample, though the enhancement is less significant compared to the 10% Ca-substituted sample. Nevertheless, electron doping presents a promising route to further enhance the wFM moment and stabilize the wFiM state predicted by our DFT calculations.

Table S13. Refined lattice parameters for $SrMnO_{2.97}$ based on laboratory XRD data collected at room temperature.

| | |
|---|---|
| *a* (Å) | 5.446(1) |
| *b* (Å) | 9.432(1) |
| *c* (Å) | 9.095(1) |
| Volume (Å$^3$) | 234.49(5) |

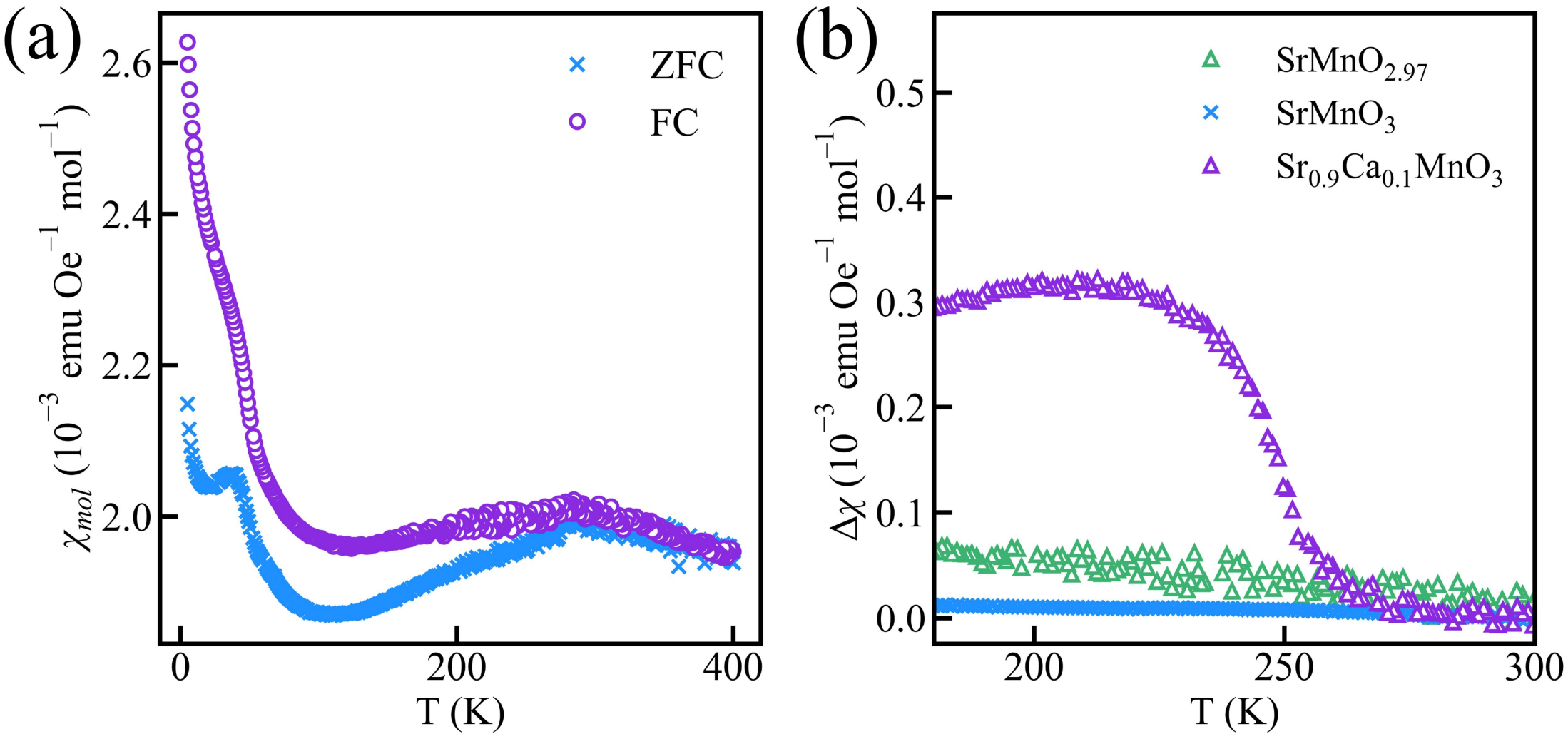

Figure S21. (a) $\chi_{\mathrm{mol}}$ for 4H-$SrMnO_{2.97}$. The peak near ~45 K is due to a $Mn_3O_4$ impurity, while a low-temperature Curie tail indicates the presence of a residual paramagnetic component. (b) Comparison of $\Delta\chi_{\mathrm{mol}}$ to $SrMnO_3$ and $Sr_{0.9}Ca_{0.1}MnO_3$, showing that the wFM component is enhanced for the reduced sample, but smaller compared to the 10% Ca-substituted sample.

Supplementary Note 8. DC resistivity measurements.

We sought to establish the electrical transport properties of our samples to assess the feasibility of performing switching experiments. Our variable-temperature DC resistivity data (Figure S22a) show that 4H-$SrMnO_3$ behaves as a semiconductor at room temperature, similar to the canonical ME perovskite $BiFeO_3$. However, on cooling to $T_N$, the resistivity rapidly becomes immeasurably large, suggesting the long-range AFM ordering is linked to an abrupt change in band gap. This is consistent with the AFM insulating ground state predicted by our DFT calculations (Supplementary Note 3). Above 300 K, the transport does not obey a simple Arrhenius dependence (Figure S22b), and neither polaron nor variable-range hopping models yielded good fits against the observed data. Close inspection of the derivative (Figure S23a) suggests this is due to a resistivity anomaly near ~350 K, but no corroborating feature of this appears in our diffraction experiments or our magnetic susceptibility measurements. This also does not appear to be related to the $Mn_3O_4$ impurity we identified in our diffraction experiments. Interestingly, we find that the anomaly disappears upon the application of a 70 kOe field (Figure S23b), at which point the transport mechanism can be modelled well with a 3D Mott variable-range hopping (VRH) mechanism (Figure S24). This suggests there is a field-induced change in charge transport, which is potentially consistent with the apparent spin-flop transition we observed in our magnetization measurements (Figure S20).

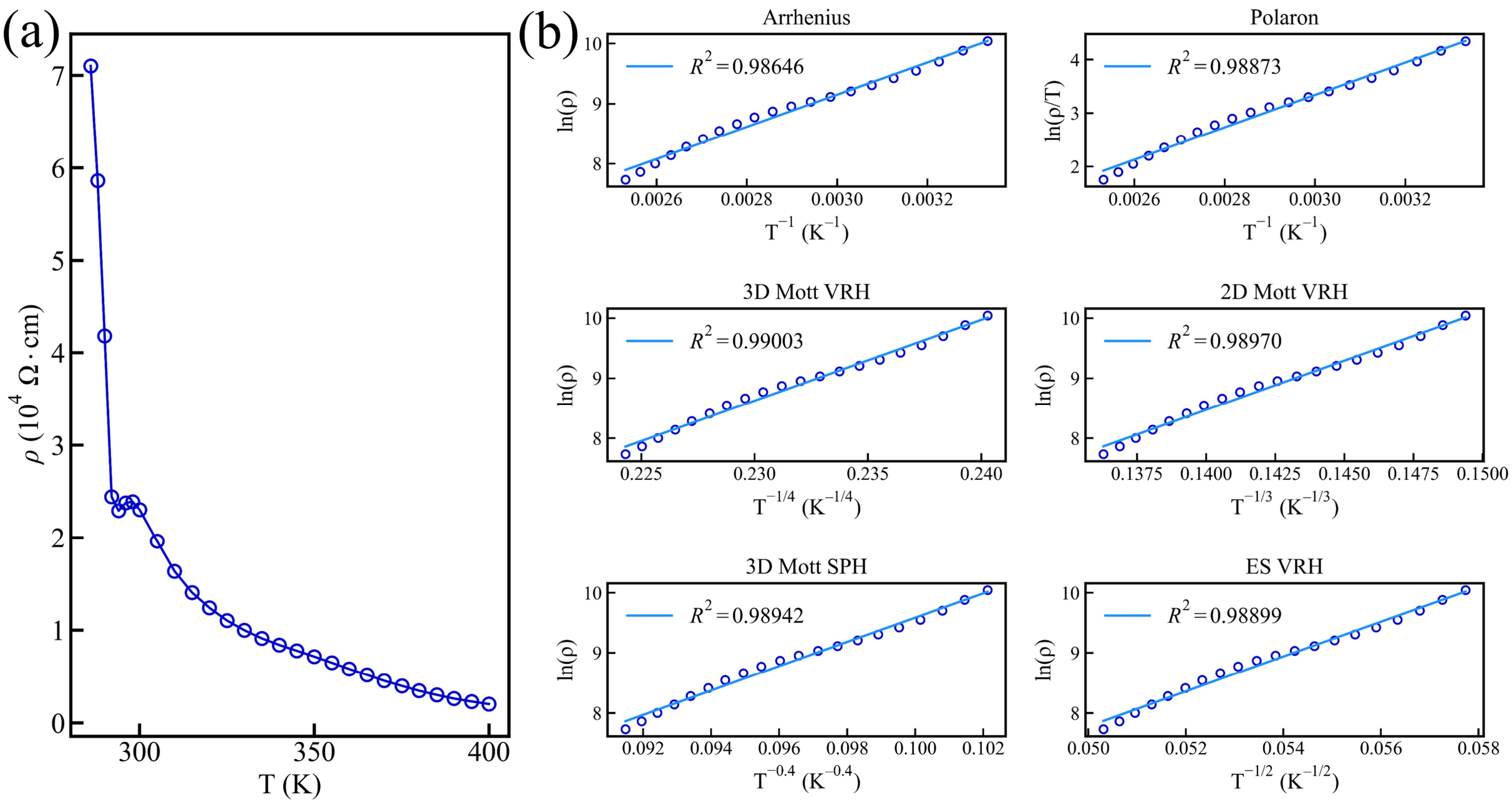

Figure S22. DC resistivity ($\rho$) measurements on 4H-$SrMnO_3$ performed in the absence of an applied magnetic field. (a) $\rho(T)$ in the range 280–400 K; $\rho$ becomes immeasurably large below 285 K. (b) Comparison of various transport models fitted in the range 300–400 K. "VRH" and "SPH" refer to variable-range hopping and spin polaron hopping, respectively, while "ES VRH" refers to Efros-Shlovskii VRH. The $R^2$ values have been overlaid for each plot for comparison.

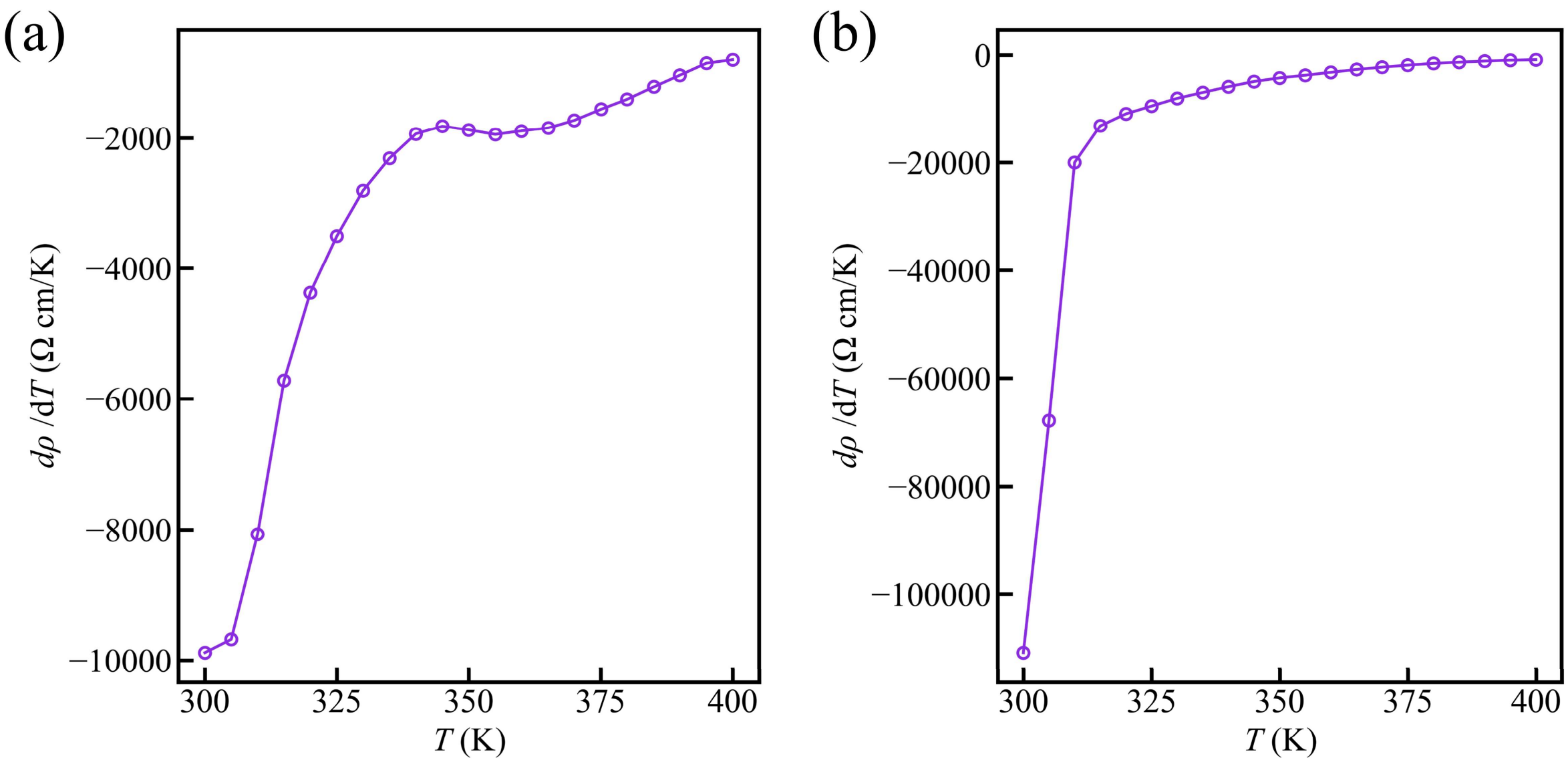

Figure S23. Derivative of $\rho$ with respect to $T$ for the data collected under (a) 0 kOe and (b) 70 kOe, respectively. In both cases, data were collected on cooling.

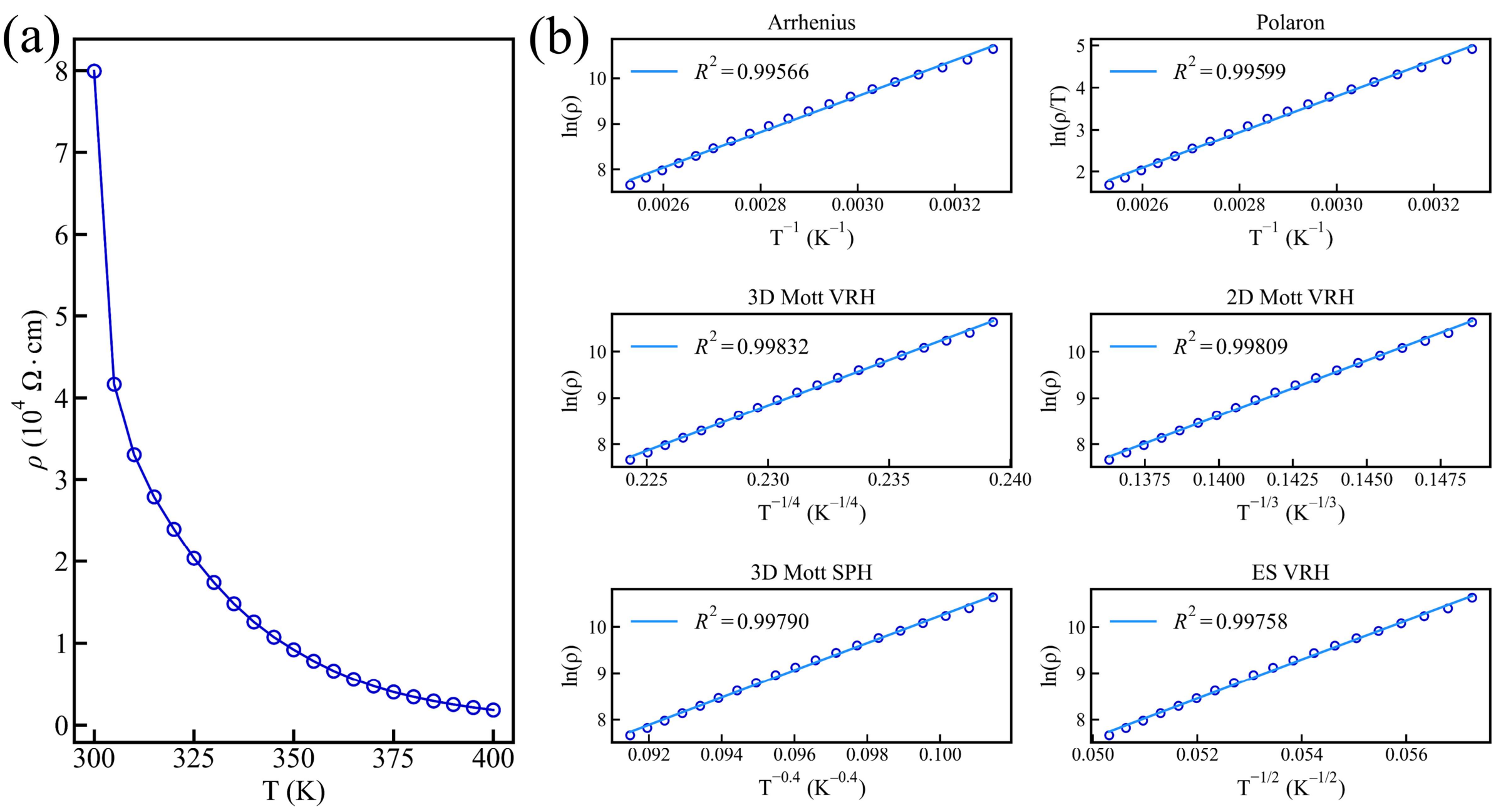

Figure S24. DC resistivity ($\rho$) measurements on 4H-$SrMnO_3$ performed with a 70 kOe field. (a) $\rho(T)$ in the range 300–400 K; $\rho$ becomes immeasurably large below 300 K. (b) Comparison of various transport models fitted in the range 300–400 K. The $R^2$ values have been overlaid on each plot for comparison.

Supplementary Note 9. Landau free energy analysis.

The nature of the coupling between the order parameters in 4H-$SrMnO_3$ can be determined by analyzing the symmetry-invariant polynomials in the Landau free energy expansion of the system. For this, we treat the $\Gamma_5^-$ (a,b) tilts, the $\Gamma_2^-$ (c) polar mode, the m$\Gamma_6^-$ (d,e) AFM component, and the m$\Gamma_6^+$ (f,g) wFM component as the salient order parameters; for convenience, we rewrite the 2D order parameters in polar form, leading to ($\rho$, $\varphi$) for $\Gamma_5^-$ and ($M_1$, $\theta_1$), ($M_2$, $\theta_2$) for m$\Gamma_6^-$ and m$\Gamma_6^+$, respectively. We also rewrite $\Gamma_2^-$ (c) as $P$, taking the magnitude of any spontaneous polarization to be proportional to the atomic displacements which transform under this irrep. For simplicity, we neglect the effects of the $\Gamma_5^+$ ferroelastic strain as its magnitude is essentially negligible in the ground state (Figure S9), and we also neglect contributions from any of the other secondary order parameters.

In our model, we consider terms up to the sixth order of the free energy expansion so as to account for the angular anisotropy of the $\Gamma_5^-$, m$\Gamma_6^-$, and m$\Gamma_6^+$ terms. The full free energy expansion according to these conditions is extremely complex, so it has been provided alongside the Supplementary Material. For the purposes of our discussion, we instead consider a minimal model which considers only the lowest-order coupling terms between the salient order parameters. Although isotropic invariants such as those found in the biquadratic terms $F \sim \gamma_1\rho^2P^2$ or $F \sim \gamma_2 M_1{}^2 M_2{}^2$ will be essential to the overall stability of the low-symmetry phases, we have purposefully omitted these terms to ensure our model only encompasses the minimum number of terms necessary to encapsulate the overall form of the energy landscape. Our minimal model is thus as follows:

$$\begin{aligned} F = \alpha_1\rho^2 + \alpha_2\rho^4 + \alpha_3\rho^6 + \alpha_4\rho^6(10 + \cos 6\varphi) \\ + \alpha_5 P^2 + \alpha_6 M_1^2 + \alpha_7 M_1^4 + \alpha_8 M_2^2 \\ + \beta_1 P\rho^3 \sin 3\varphi + \beta_2 P M_1 M_2 \sin\Delta + \beta_3 \rho M_1 M_2 \sin\Theta \end{aligned}$$

where $\Delta = \theta_1 - \theta_2$ and $\Theta = \theta_1 + \theta_2 - \varphi + \pi/6$. Given that the $\Gamma_5^-$ and m$\Gamma_6^-$ modes are unstable with respect to the parent paramagnetic structure, we take $\alpha_1$, $\alpha_6 < 0$ and include their higher-order terms to maintain boundedness. We also include the sixth-order terms in $\rho$ to account for the anisotropy of the tilts, which dictates the relative stability of the $C222_1$ ($\alpha_4 < 0$, $\varphi = n\pi/3$) and $Cmc2_1$ ($\alpha_4 > 0$, $\varphi = (2n + 1)\pi/6$) structural phases. Analogous anisotropy terms are also present for the m$\Gamma_6^-$ and m$\Gamma_6^+$ orders, but we assume their contribution to the overall energy landscape is negligible compared to the structural order parameters.

First, we consider the coupling between the tilts and the polarization, which is of the form:

$$F \sim \beta_1 P\rho^3 \sin 3\varphi$$

The linear coupling in $P$ ensures it may adopt either a positive or negative sign so as to minimize $F$, hence this can always favor the $Cmc2_1$ phase ($\varphi = (2n+1)\pi/6$) regardless of the sign of the $\beta_1$ coefficient.

Symmetry-related $Cmc2_1$ domains are related by $\Delta\varphi = \pm n\pi/3$ (see Figure S6); for odd values of $n$, the direction of $P$ necessarily flips by 180° to maintain invariance overall. This represents the most likely basis for polarization reversal in 4H-$SrMnO_3$, similar to what has been proposed in hexagonal rare-earth manganites such as $YMnO_3$ [17].

Next, we consider the trilinear coupling between the tilts, the AFM order, and the wFM order, which takes the form:

$$F \sim \beta_2 P M_1 M_2 \sin\Delta, \qquad \Delta = \theta_1 - \theta_2$$

Again, the linear nature of the coupling in $P$ ensures it may adopt either a positive or negative sign so as to minimize $F$. The transformation $P \rightarrow -P$ necessarily induces $\Delta \rightarrow \Delta \pm \pi$ to maintain invariance. This can occur via the rotation of either $\theta_1$ or $\theta_2$ by 180°, but since the switching barriers are likely to be much higher for $\theta_1$, reorienting $\theta_2$ is far more likely. Nevertheless, this simple picture neglects the role of the coupling to the tilts, which must also reorient as part of any proposed switching mechanism. To consider this on an explicit basis, we also examine the additional trilinear invariant:

$$F = \beta_3 \rho M_1 M_2 \sin\Theta, \qquad \Theta = \theta_1 + \theta_2 - \varphi + \frac{\pi}{6}$$

Here, the 2D anisotropies of the tilts and the AFM/wFM components have been absorbed into the single phase angle $\Theta$. Minimizing $F$ with respect to $\Theta$ leads to the condition $\theta_1 + \theta_2 = \varphi - \pi/6$, showing how the orientation of the magnetic order parameters depends upon the orientation of the $\Gamma_5^-$ tilts. A reversal in $P$ may effect the transformation $\varphi \rightarrow \varphi \pm \pi$. Given $F$ is also minimized for $\Delta = \pm\pi/2$, this triggers a 180° reorientation in either $\theta_1$ or $\theta_2$. Again, $\theta_2$ is much more likely to reorient based on the anticipated switching barriers for the AFM and wFM moments. Verifying the feasibility of this proposed switching mechanism would require a bespoke computational study assessing the relative switching barriers associated with $\Delta\varphi = n\pi/3$, which lies far beyond the scope of the present work. Nevertheless, 4H-$SrMnO_3$ possesses the capacity for ME switching, and further work should seek to explore different switching pathways to facilitate its experimental realization.

Supplementary References